\documentstyle[12pt,epsfig]{article}

\advance\textheight by \topskip
\newcommand{\be}{\begin{equation}}
\newcommand{\ee}{\end{equation}}
\newcommand{\br}{\begin{eqnarray}}
\newcommand{\bea}{\begin{eqnarray}}
\newcommand{\beanon}{\begin{eqnarray*}}
\newcommand{\er}{\end{eqnarray}}
\newcommand{\eea}{\end{eqnarray}}
\newcommand{\eeanon}{\end{eqnarray*}}
\newcommand{\ba}{\begin{array}}
\newcommand{\ea}{\end{array}}
\newcommand{\bi}{\begin{itemize}}
\newcommand{\ei}{\end{itemize}}
\newcommand{\bn}{\begin{enumerate}}
\newcommand{\en}{\end{enumerate}}
\newcommand{\bc}{\begin{center}}
\newcommand{\ec}{\end{center}}

\newcommand{\ar}{\rightarrow}

\newcommand{\Dir}{\kern -6.4pt\Big{/}}
\newcommand{\Dirin}{\kern -10.4pt\Big{/}\kern 4.4pt}
\newcommand{\DDir}{\kern -7.6pt\Big{/}}
\newcommand{\DGir}{\kern -6.0pt\Big{/}}
\newcommand{\eett}{$e^+e^-\rightarrow t\bar t $}
\newcommand{\eeZH}{$e^+e^-\rightarrow ZH $}
\newcommand{\eeAH}{$e^+e^-\rightarrow AH $}
\newcommand{\eeWWh}{$e^+e^-\rightarrow hW^+W^-$}
\newcommand{\eebbww}{$e^+e^-\rightarrow b\bar b W^+W^-$}
\newcommand{\ttbbww}{$t\bar t\rightarrow b\bar b W^+W^-$}
\newcommand{\eettbbww}{$e^+e^-\rightarrow t\bar t\rightarrow b\bar b W^+W^-$}

\newcommand{\bbww}{$b\bar b W^+W^-$}
\newcommand{\eeb}{$ e^+e^-$}

\newcommand{\ttb}{$ t\bar t$}
\def\Ord{\buildrel{\scriptscriptstyle <}\over{\scriptscriptstyle\sim}}
\def\OOrd{\buildrel{\scriptscriptstyle >}\over{\scriptscriptstyle\sim}}
\def\sm{\ifmmode{{\cal {SM}}}\else{${\cal {SM}}$}\fi}
\def\mssm{\ifmmode{{\cal {MSSM}}}\else{${\cal {MSSM}}$}\fi}
\def\MH{\ifmmode{{M_{H}}}\else{${M_{H}}$}\fi}
\def\Mh{\ifmmode{{M_{h}}}\else{${M_{h}}$}\fi}
\def\MA{\ifmmode{{M_{A}}}\else{${M_{A}}$}\fi}
\def\MHpm{\ifmmode{{M_{H^\pm}}}\else{${M_{H^\pm}}$}\fi}
\def\tb{\ifmmode{\tan\beta}\else{$\tan\beta$}\fi}
\def\ctb{\ifmmode{\cot\beta}\else{$\cot\beta$}\fi}
\def\ta{\ifmmode{\tan\alpha}\else{$\tan\alpha$}\fi}
\def\cta{\ifmmode{\cot\alpha}\else{$\cot\alpha$}\fi}
\def\tba{\ifmmode{\tan\beta=1.5}\else{$\tan\beta=1.5$}\fi}
\def\tbb{\ifmmode{\tan\beta=30.}\else{$\tan\beta=30.$}\fi}
\def\cab{\ifmmode{c_{\alpha\beta}}\else{$c_{\alpha\beta}$}\fi}
\def\sab{\ifmmode{s_{\alpha\beta}}\else{$s_{\alpha\beta}$}\fi}
\def\cba{\ifmmode{c_{\beta\alpha}}\else{$c_{\beta\alpha}$}\fi}
\def\sba{\ifmmode{s_{\beta\alpha}}\else{$s_{\beta\alpha}$}\fi}
\def\ca{\ifmmode{c_{\alpha}}\else{$c_{\alpha}$}\fi}
\def\sa{\ifmmode{s_{\alpha}}\else{$s_{\alpha}$}\fi}
\def\cb{\ifmmode{c_{\beta}}\else{$c_{\beta}$}\fi}
\def\sb{\ifmmode{s_{\beta}}\else{$s_{\beta}$}\fi}
\def\pl #1 #2 #3 {{\it Phys.~Lett.} {\bf#1} (#2) #3}
\def\np #1 #2 #3 {{\it Nucl.~Phys.} {\bf#1} (#2) #3}
\def\zp #1 #2 #3 {{\it Z.~Phys.} {\bf#1} (#2) #3}
\def\pr #1 #2 #3 {{\it Phys.~Rev.} {\bf#1} (#2) #3}
\def\prep #1 #2 #3 {{\it Phys.~Rep.} {\bf#1} (#2) #3}
\def\prl #1 #2 #3 {{\it Phys.~Rev.~Lett.} {\bf#1} (#2) #3}
\def\mpl #1 #2 #3 {{\it Mod.~Phys.~Lett.} {\bf#1} (#2) #3}
\def\rmp #1 #2 #3 {{\it Rev. Mod. Phys.} {\bf#1} (#2) #3}
\def\xx #1 #2 #3 {{\bf#1}, (#2) #3}
\def\preprint{{\it preprint}}
\begin{document}
\tolerance=100000
\thispagestyle{empty}
\setcounter{page}{0}

\begin{flushright}
{\large Cavendish--HEP--96/3}\\
{\large DFTT 19/96}\\ 
{\rm March 1996\hspace*{.5 truecm}}\\ 
\end{flushright}

\vspace*{\fill}

\begin{center}
{\Large \bf 
The process $e^+e^-\rightarrow b\bar bW^+W^-$ \\
at the Next Linear Collider\\
in the Minimal Supersymmetric Standard Model}\\[2.cm]
{\large Stefano 
Moretti\footnote{E-mails: 
moretti@hep.phy.cam.ac.uk,moretti@to.infn.it.}}\\[.3 cm]
{\it Cavendish Laboratory,
University of Cambridge,}\\
{\it Madingley Road,
Cambridge, CB3 0HE, United Kingdom.}\\[0.5cm]
{\it Dipartimento di Fisica Teorica, Universit\`a di Torino, Italy}\\
{\it and INFN, Sezione di Torino}\\
{\it Via Pietro Giuria 1, 10125 Torino, Italy.}\\[1cm]
\end{center}

\vspace*{\fill}

\begin{abstract}
{\normalsize
\noindent
The complete matrix element for $e^+e^-\ar b\bar b W^+W^-$
is computed at tree-level within the Minimal Supersymmetric
Standard Model. 
Rates of interest to phenomenological analyses at the Next Linear
Collider are given. In particular, we study:
\begin{itemize}
\item $t\bar t$ production and decay $t\bar t\ar (bW^+)(\bar bW^-)$;
\item $ZH$ production followed by $Z\ar b\bar b$ and $H\ar W^+W^-$;
\item $AH$ production followed by $A\ar b\bar b$ and $H\ar W^+W^-$;
\item $hW^+W^-$ production followed by $h\ar b\bar b$. 
\end{itemize}
Top and Higgs finite width effects are included,
as well as all those of the irreducible backgrounds.}
\end{abstract}

\vspace*{\fill}
\newpage

\section*{1. Introduction} 

By the time that the Next Linear Collider (NLC) \cite{ee500} will be operating,
both the exact value of
the top 
quark mass and the structure of the Higgs sector of the electroweak (EW)
interactions will be already known, thanks to the combined
action of the Tevatron \cite{Teva}, of
LEP~II \cite{lep2w} and of the LHC \cite{CMS,ATLAS}.   
It is however clear that detailed studies of both top quark and Higgs
boson properties will have to wait the advent of an \eeb\ linear machine. 

In two previous papers \cite{tt,ZH} detailed studies of the process
\eebbww\ at centre-of-mass (CM) energies typical of the
NLC were presented. Those analyses were concerning 
the Standard Model (\sm). In that framework, 
the importance of the process \eebbww\ is
evident if one considers that it represents a signature to top production
in $t\bar t$ pairs as well as to that of the \sm\ Higgs boson $\phi$ 
in the $Z\phi$ channel. In fact, on the one hand, top pairs produced
via the process \eett\ decay 
through $t\bar t\ar (bW^+)(\bar bW^-)$ whereas, on the
other hand, the channel $Z\phi\ar (b\bar b)(W^+W^-)$ might well be one of best
ways to detect a heavy Higgs, thanks to the expected performances of the
vertex detectors in triggering the $Z$ boson \cite{ideal}\footnote{The mode 
$Z\ar b\bar b$ has a branching ratio (BR)
about five times larger than the BRs into $\mu^+\mu^-$
or $e^+e^-$ and it is equally free from backgrounds coming from $W$ decays.}.
From those studies, the importance of top finite width effects and of those 
due to the non-resonant
background in $b\bar bW^+W^-$ events clearly came out, together with 
positive prospects of Higgs detection.

It is the purpose of this report to extend those analyses to the case 
of the Minimal Supersymmetric Standard Model (\mssm).
In this context, the reaction \eebbww\ is important in at least four
respects.
First and second, like in the \sm, it allows studies of top pair production and 
decay $t\bar t\ar (bW^+)(\bar bW^-)$ as well as of the Higgs channel 
$ZH\ar (b\bar b)(W^+W^-)$.
Third and fourth, it also allows one to analyse Higgs production in 
the channels $e^+e^-\ar AH\ar (b\bar b)(W^+W^-)$ and 
$e^+e^-\ar hW^+W^-\ar (b\bar b)W^+W^-$.

In particular, we notice that the \sm-like channel $t\ar bW^\pm$ is 
the dominant top decay mechanism over a large
part of the \mssm\ parameter space, especially if the mass of the charged
Higgs boson $H^\pm$ is comparable to $m_t$, such
that the decay channel $t\ar bH^\pm$ is strongly 
suppressed by the available phase 
space\footnote{Throughout this paper
we assume that the mass scale of the Supersymmetric partners
of the ordinary elementary particles is well beyond the energy reach
of the NLC, such that they cannot be produced at this machine.
In particular, we neglect here considering the Supersymmetric decay
$t\ar \tilde{t}\tilde{\chi}_1^0$, where $\tilde{t}$ represents the
stop and $\tilde{\chi}_1^0$ refers to the lightest
neutralino, as well as other possible \mssm\ modes. For a 
recent review of these latter, see Ref.~\cite{Guasch}.} \cite{bern}. 
Conversely, when this is not the case,
finite width effects should 
be more important in the \mssm\ \cite{ref30b,ref27}, 
as for $\MHpm<m_t-m_b$ (i.e., small
values of $\MA$) one gets $\Gamma_t^{\mssm}>\Gamma_t^{\sm}$.   
Furthermore, the process $e^+e^-\ar ZH$ is nothing else than 
counterpart of the \sm\ Higgs bremsstrahlung mechanism,
where the heaviest of the neutral scalar Higgses of the \mssm, $H$, 
plays the r\^ole of $\phi$ \cite{proceed}.
Finally, the decays $h,A\ar b\bar b$ (of the light scalar and of the 
pseudoscalar Higgs bosons, respectively) 
and $H\ar W^+W^-$ can be the dominant ones of these 
particles  over a sizable portion of the plane $(M_A,\tan\beta)$, with 
the rates of \eeAH\ being comparable to those of \eeZH\ \cite{DKZ}
and the cross sections for \eeWWh\ being possibly of a few picobarns
for $\Mh\Ord100$ GeV (the increased $hb\bar b$ coupling compensating
the reduced $hW^+W^-$ one, with respect to the corresponding \sm\ values).

We further remind the reader that the four processes \eett,
\eeZH, \eeAH\ and \eeWWh\ cannot be unambiguously separated, and studied
independently one from the others.
Therefore any of them constitutes an irreducible background  
to the other three in the $b\bar bW^+W^-$ channel, and such interplay 
must be carefully
taken into account when studying full \eebbww\ events. As we are
here computing the complete Matrix Element (ME) of such a process,
also non-resonant background effects will be present. In this respect,
diagrams involving the lightest \mssm\ neutral Higgs $h$, 
in which this is produced either far below
threshold (in the splitting $h^*\ar W^+W^-$) or in association with a
off-shell $Z$ (via $Z^*\ar W^+W^-$), must be
regarded as background.

The plan of the paper is as follows. In the next Section we describe
the calculation we have performed; in Section 3 we present
our results, whereas the conclusions are given in Section 4.

\section*{2. Calculation} 

In the \mssm\ the process
\be\label{eebbww}
e^+e^-\ar b\bar b W^+ W^-
\ee
proceeds at tree-level through 75 Feynman diagrams, which can
be conveniently grouped in 28 different topologies. 
Of these latter, 19 correspond to actual \sm\ graphs, 8 to Higgs processes
with ${\cal{SM}}$-like structure (in which a superposition of the $h$ and $H$
bosons replaces the $\phi$ scalar) and 2 to typical \mssm\ graphs involving
a vertex
with two (different) Higgs scalars and one vector boson. 
The first 27 topologies can be found in Refs.~\cite{tt,ZH},
whereas the last one is depicted in Fig.~1. 

To compute the amplitude squared of process (\ref{eebbww}) we have
used two different spinor formalisms, described in 
Refs.~\cite{gattoevolpe,HZ}. 
The results produced by the two corresponding {\tt FORTRAN} codes agree
within 10 significative digits in {\tt REAL*8} precision. Moreover,
the amplitude produced with the method of Ref.~\cite{HZ} has been
tested for gauge invariance, as it has been
implemented in three different  gauges (Unitary, Feynman and Landau). 
For the technical details of the 
numerical evaluation of the ME and of its integration
over the phase space we refer to Ref.~\cite{tt}, as we 
have adopted here the same procedures.

For the parameters which are in common to the \sm\ and to the \mssm\
we have used the following numerical values:
$M_{Z}=91.175$ GeV, $\Gamma_{Z}=2.5$ GeV,
$M_{W^\pm}=M_{Z}\cos\theta_W\approx80$ GeV, 
$\Gamma_{W^\pm}=2.2$ GeV and $m_b=4.25$ GeV.
For simplicity, we have kept the final state bosons $W^\pm$ on-shell
in the computation. 
The expressions for the \mssm\ Higgs mass relations and couplings that
we have used are the same that have been 
summarised in Ref.~\cite{ioejames} (see also references
therein).
The widths of the \mssm\ Higgs bosons have been generated by using
the {\tt FORTRAN} program described there.
In particular, as we have adopted running $b$-masses in evaluating
$\Gamma_H,\Gamma_h$ and $\Gamma_A$, in order to be consistent, we have used 
the same $m_b(Q^2=M_\Phi^2)$ mass also in the $\Phi b\bar b$ vertices
of the production processes considered here, where $\Phi=H,h$ and $A$.
The values adopted for the strong coupling constant $\alpha_s$ are those
at two-loops, for $n_f=5$ active flavours and 
$\Lambda_{QCD}^{n_f=5}=0.13$ GeV, at the scale $Q=\sqrt s$,
in the $\overline{\mathrm{MS}}$
renormalisation scheme and with $\alpha_s(M_Z)=0.112$. 

As it is clearly impractical to cover all regions of the
\mssm\ parameter space, we have chosen here,
as representative for
$\tan\beta$, the two extreme values
1.5 and 30., whereas $M_{A}$
spans the range 50 to 350 GeV, that is, between the experimental lower
limit \cite{Alimit} and the region  
where the $t\bar t$ channels of \mssm\ neutral Higgs bosons start dominating
the decay phenomenology over a substantial part of the $(\MA,\tan\beta)$ 
plane \cite{ioejames}. 

The top width $\Gamma_t$ has been evaluated according to the 
formulae given in Ref.~\cite{widthtopSM1} (see also Ref.~\cite{widthtopSM2}), 
corrected by means of
the expressions of Ref.~\cite{widthtopMSSM}, 
to account for the \mssm\ decay $t\ar bH^\pm$.
In order to obtain results in Narrow Width Approximation (NWA)
for the top, we have written the heavy quark propagator as
\be\label{propagator}
\frac{ p\Dir  + m_t}{p^2-m_t^2+im_t\Gamma}
\left( \frac{\Gamma}{\Gamma_{t}}\right)^{\frac{1}{2}}.
\ee
In this way, for $\Gamma = \Gamma_{t}$
the standard expression is recovered, whereas for
$\Gamma \ar 0$ one is able to correctly reproduce the rates for \eett\ times
the (squared) branching ratio $[BR(t\ar bW^\pm)]^2$. Numerically, we have used
$\Gamma=10^{-5}$ in NWA.

For the discussion of the results we have assumed the following values
for the top mass: $m_t=170,172$ and 174 GeV \cite{CDFtop} and
$m_t=195,197$ and 199 GeV \cite{D0top}. In correspondence, we have
taken as CM energy $\sqrt s=350$ and 400 GeV. This has been done in order 
to perform studies of top-antitop production at threshold, 
according to the values of $m_t$ measured by CDF and D0, 
respectively\footnote{When this work was almost completed
the Fermilab Collaborations have both announced new measurements of $m_t$ 
\cite{newtop}, which seem to
shift the top mass towards the lower part of the $m_t$ spectrum considered
here. However, as also the new values suffer from rather large uncertainties
(such that by summing statistics and systematics the 
experimental error band would include the most part of top masses 
that we have chosen),  
we decided to maintain in the present paper 
also the part of studies devoted to the case $m_t\approx199$ GeV.}.
As top measurements will be certainly performed also far above
threshold, we have produced results for the combination
$\sqrt s=500$ GeV and $m_t=174$ and 199 GeV too.
The total integrated luminosity assumed for the NLC throughout this
paper is ${\cal L}=10$ fb$^{-1}$. Please also note that
in all Tables and Figures the branching ratios of the $W$'s
have not been included. 

Finally, we are aware that a great number of higher-order results 
have been presented concerning $t\bar t$ production and decay (for a review,
see Ref.~\cite{reviewtop}).
For example, QCD correction to the Born cross section up to first order
in $\alpha_s$ have been calculated in Ref.~\cite{orange54}.  
These corrections lead to the familiar Coulomb enhancement
at threshold (this also occurs
in QED) \cite{vak}. The QED-like binding potential between the two
top quarks at threshold decreases however rather rapidly \cite{ref9}, such
that for $\Delta E\equiv\sqrt s-2m_t\OOrd2$ GeV the two fermions behave
as free particles and ordinary perturbation theory can be applied
(our approach will be indeed to concentrate our attention in this region,
referring for the complementary case $\Delta E\Ord2$ GeV to more refined
studies \cite{vak,ref9,thr}). EW
corrections have been studied
in Ref.~\cite{orange58,newcorr}. These latter are generally negative. 
Both of them account for an ${\cal O}(10\%)$ effect  (for
$m_t>170$ GeV). We do not include such corrections here, for two reasons.
On the one hand, for consistency, as these are known only 
for the process $e^+e^-\ar t\bar t\ar b\bar b W^+W^-$ but not
for the non-resonant background (which is computed here at tree-level 
only). On the other hand, because the typical spikes  at threshold due to
Coulomb effects are largely smeared out in the excitation curve if
the top mass is large, see Ref.~\cite{bern,Bagliesi}.

Also concerning the \mssm\ Higgs processes 
\eeZH\ and \eeAH\ higher-order results have
been reported in literature \cite{Poland,Dabel}. 
These are EW corrections, which are generally under control, if the scale
of the SUSY partners is set in the TeV region, as we assume here
\cite{genuine,expre}.
Again reasons of consistency have induced us to treat Higgs signals
and corresponding background in the same way 
also in this case (i.e., at leading order). The Higgs reaction
\eeWWh\ has been
studied within the \sm\ in Ref.~\cite{WWh}, and no higher-order results are
known to date. 

Beamsstrahlung and Linac energy spread effects \cite{ISR}
have been systematically neglected here, for two reasons.
On the one hand, effects due to synchrotron radiation emitted by one of 
the colliding
bunches in the field of the opposite one 
as well as the intrinsic (to any collider)
energy distribution of the beams before annihilation 
necessarily need, in order
to be quantified, the knowledge of the technical details
of the collider design and can be realistically estimated only through
Monte Carlo simulations. On the other hand, it has been shown that, 
for narrow beam designs, beamsstrahlung
affects the cross section much less than the Initial State Radiation
(ISR) \cite{ISR}, such that in
phenomenological analyses one can consistently 
deal with bremsstrahlung radiation only.  

Therefore, we will devote some space to discussing the effect of
the ISR emitted from electron/positron lines \cite{ISR}.
In particular, we expect the signal $t\bar t$ to be quite 
sensitive at threshold \cite{bern}. In this respect, we confine ourselves
to some illustrative examples in the presence of ISR, more than implementing
this latter in all cases. Both because ISR effects are expected to be
qualitatively the same regardless of 
the actual values of the various parameters
($m_t,\Gamma_t,\MA,\tan\beta$, etc.) and because its inclusion would
significantly enhance the CPU time of the runs. In general, however, 
ISR effects are straightforward to include \cite{structure,Nicro}. 
The main effect of the ISR is to lower the effective
CM energy available in the main process, thus ultimately
reducing(enhancing)
the total cross sections which
increase(decrease) at larger CM energies.
Furthermore, ISR  also leads to a smearing of the
differential distributions \cite{BCDKZ,Orange3}.

In the very end, in order to perform the foreseen high precision measurements
of the top parameters ($m_t,\Gamma_t,\tau_t$, etc.) and to
disentangle Higgs signals at the NLC, a 
complete phenomenological analysis has to include in full
all the above corrections \cite{ee500}. This is clearly beyond the 
scope of this study. 
For the moment, we are mainly concerned with the fact that other aspects so far
either ignored (irreducible background in non-top-antitop events) 
or only approximated in the existing
literature (finite width
and spin correlation effects in the top production and decay)
 could be important, such that they
must be properly taken into account for a correct analysis, and
to check if new Higgs discovery channels could be of experimental 
interest.
To realise whether all of this is true or not, and when, 
is the final goal of the present paper.

\section*{3. Results}

In carrying out our analysis we closely follow the approach of
Refs.~\cite{tt,ZH}, and the phenomenological studies reported in various
instances  
in Ref.~\cite{ee500}. 
For future reference, we show in Figs.~2a--b
the \mssm\ mass relations between, on the one hand, the $H^\pm,h$
and $H$ Higgs scalars and, on the other hand,
the $A$ pseudoscalar,  as well as 
the partial decay widths of the top quark in the two \mssm\
channels $t\ar bH^\pm$
and $t\ar bW^\pm$, for the mentioned combinations of 
$\tan\beta$ and $\MA$. 

Throughout this paper,
of the two $W$ bosons in the 
\bbww\ final state, one is assumed to decay hadronically, 
$W\ar \mathrm{jj}$, and 
the other leptonically, $W\ar \ell\bar\nu_\ell$ with $\ell=e,\mu$. 
This semi-leptonic (or semi-hadronic) final state
has a few advantages with respect to the case of two hadronic $W$ decays:
it has a simpler topology inside the detectors 
and thus it is easier to reconstruct;
it also allows one to get rid of complications due to the combinatorial
in case of a six jet final state. Moreover, like the pure hadronic final state
its kinematics is fully constrained (once the missing momentum 
is assigned to the neutrino). Finally, two leptonic $W$ decays
would lead to a double disadvantage: first, a very much reduced statistics
and, second, problems in reconstructing invariant mass spectra because
of the two neutrinos.

We present our results by treating the top and Higgs production mechanisms
in two different Subsections. 

\subsection*{3.1 The process $e^+e^-\ar t\bar t\ar (bW^+)(\bar bW^-)$}

The selection strategies of top-antitop signals we have 
considered are (see Ref.~\cite{Bagliesi}):
\begin{itemize}
\item to perform a scan in $\sqrt s$;
\item to study the $W$ momentum spectrum;
\item to reconstruct the invariant mass of the three-jet system
$t\ar bW^\pm$.
\end{itemize}
The first two methods are normally used at threshold ($\sqrt s\approx 2m_t$), 
whereas the last one has been considered for studies far above that 
($\sqrt s\gg 2m_t$). 

In the first method one spans the CM energy in a region
of approximately 10 GeV around the real top-antitop threshold, in order
to reconstruct the $t\bar t$-excitation curve. By using the fit procedures
described in Refs.~\cite{bern,Bagliesi} one should be able to measure the
top mass as well as to perform a large 
number of measurements of fundamental quantities, such as the
strong coupling constant \cite{vak,ref9}. 
Moreover, if $m_t$ and $\alpha_s$ can be measured independently and
high experimental precision can be achieved,
it may be possible to extract from the data also
the values of $\Gamma_t$ and/or to obtain limits on the mass(es) 
$M_{\phi(\Phi)}$
and the top Yukawa coupling(s) $\lambda_{\phi(\Phi)}$
of the Higgs boson(s) of the underlying theory 
\cite{bern}. In order to perform a careful analysis, one should
include in the end \cite{bern}: complete EW corrections
to $t\bar t$ production \cite{ref7}, higher-order QCD corrections
(including `bound state' and `short distance' effects) and 
beam-related phenomena \cite{Bagliesi,ISR}. 

In the second approach, one exploits the fact that, with the CM energy 
constraint
and after assigning the missing energy and momentum to the neutrino,
the $W$ boson four-momentum $p_W$ can be easily reconstructed by adding
those of the lepton and of the neutrino, and that the width of the
corresponding differential spectrum is sensitive to the top quark mass
\cite{Bagliesi}. Uncertainties due to beam-related effects
(ISR, beamsstrahlung) are expected to be generally small,
whereas those due to the `Fermi motion' of the top quarks should be
considered in detail in a full simulation \cite{Bagliesi}.

In the third case, one determines $m_t$ from the 
two three-jet invariant mass 
distributions ($M_{bW\ar\mathrm{3jet}}$) that can be reconstructed
from the $b\bar bW^+W^-\ar b\bar b(\mathrm{jj})(\ell\bar\nu_\ell)$ 
final state,
after appropriate selection cuts to reduce the non-top 
background \cite{Bagliesi}. In fact, in \bbww\ events
in which one $W$ boson decays hadronically (let us say the $W^+$)
and the other leptonically, there are two possible three-jet
combinations among the $\mathrm{jj}$ pair reconstructing the $W$ and
the two $b$'s\footnote{This is true regardless to the fact
that one uses or not $b$-tagging procedures in recognising jets from 
$b$-quarks.}: the `right' one ($W^+b$), which peaks at $m_t$, and the
`wrong' one ($W^+\bar b)$, which gives a rather broad and flat 
distribution.
In this case, systematic errors (which cause a `top mass shift')
are expected from detector and beam-related effects, as well as 
from top quark fragmentation. 
However, such a shift can be predicted to high precision \cite{Bagliesi}. 

The reason why the first two strategies are usually preferred 
to the third one at threshold 
is that for $\sqrt s\approx2m_t$ the top quarks recoil
with rather low velocity, such that in the resulting
spherical final state the fragments of the top and antitop
quarks are mixed up and it is difficult to isolate individual jets 
and to reconstruct $m_t$ from invariant mass distributions, whereas 
at higher energies the jets from 
$t\ar bW^\pm\ar b(\mathrm{jj})$ 
decays carry a momentum of a few tens of GeV, which
makes it possible to individually recognise them by simply using one of the 
usual jet finding algorithms \cite{4guys}.

The impact of using a full calculation of \eebbww\ events, which includes
all the irreducible background, 
all the finite $\Gamma_t$ effects as well as all spin correlations between
the
two top decays in the `canonical' way, 
is clear. In fact, in the literature, on the one
hand, a complete calculation
of non-\ttb-resonant events in \bbww\ final states
does not exist to date (within the \mssm) and, on the other hand, 
it is quite common (especially
when one 
has an `analytical approach' in mind, see formulae in Ref.~\cite{bern})
to split the process \eettbbww\ by separating the \eett\ production process
from the two \ttbbww\ decays, considered independently from each other,
in the spirit of the so-called `Narrow Width Approximation'\footnote{Although
width and spin correlations effects have been introduced in
various instances in previous
analyses (see Ref.~\cite{bern}, and references therein).}. 
In particular, with respect to the top search 
strategies at threshold, we stress two aspects. First, a 
correct normalisation of the  
`underlying' cross section for \bbww\ events not proceeding via
\eett\ 
is needed if one wants to disentangle subtle effects due to 
$\alpha_s$, to $\Gamma_t$ and to higher-order contributions
of Higgs boson(s), as these can be of the
same order as those due to the single-top, the $Z\ar b\bar b$ 
\cite{noi5} and the Higgs \cite{tt,ZH} 
production mechanisms in \bbww\ events.  
Second, we remind the reader that both the non-\ttb-resonant 
background and the finite value of $\Gamma_t$ will contribute to smearing out
the edges of the $p_W$ spectrum.
Concerning the three-jet invariant mass analysis, one
has to remember that \eebbww\ events that do not proceed
via \eett\ tend to enhance the 
contribution due to the `wrong' $bW$ combination, thus reducing the final
value of the signal-to-background ratios obtained from the $M_{bW}$
spectrum\footnote{In addition, although we will not treat  this issue here,
we notice that a correct inclusion of spin effects between the two top
decays is also desirable if one considers that studies of angular correlations
of the $t\bar t$ decay products are planned at the NLC (the top quark decays
before hadronising, such that its original spin polarisation is not washed 
out by multiple 
QCD soft radiation), in order to assess or disprove the presence
pf $CP$-violating effects (due to possible `New Physics') 
\cite{ref81-85,ref38,ref78}.}.

In Tabs.~I--II we present various 
cross sections for the three subprocesses
\begin{enumerate}
\item $e^+e^-\ar t\bar t$ (Narrow Width Approximation, null top width);
\item $e^+e^-\ar t\bar t\ar b\bar b W^+W^-$ (production and decay diagram
only, finite top width);
\item \eebbww\ (all diagrams at tree-level, finite top width).
\end{enumerate}
The combinations of energy and top masses are
$\sqrt s=350$ GeV and $m_t=170,172,174$ GeV (Tab.~I);
$\sqrt s=400$ GeV and $m_t=195,197,199$ GeV (Tab.~II).
In the upper part of the tables we present rates when no cut is implemented,
whereas in the lower part the constraint 
$M_{\mathrm{had}}>200$ GeV
is applied \cite{Bagliesi}, where $M_{\mathrm{had}}$ represents the invariant 
mass of the system $b\bar bW^+$, if $W^+$ is the gauge boson decaying
hadronically. The first line refers to \tba\ whereas the second corresponds
to numbers obtained by using \tbb\  
For each process and combination of masses and energies four cross sections
are given, corresponding to
$\MA=60,140,220,300$ GeV (without and inside parentheses, brackets
and braces, respectively).
As can be seen from Fig.~2b, the partial top width into $bH^\pm$ is zero
for $\MA\OOrd180$ GeV (both for \tba\ and \tbb), such that rates
for the processes $t\bar t$ and \ttbbww\ do not change in the above
range of $\MA$.

We have displayed the cross sections of processes
1.--3. in three different columns, and this
allows one to appreciate the effect of the finite width $\Gamma_t$
and of the irreducible background separately.
As already observed in the case of the \sm, rates for the process
\eett\ (first column) are larger than those for \eettbbww\ (second column),
both at $\sqrt s=350$ GeV and $\sqrt s=400$ GeV. 
The effect is clearly due to a value of $\Gamma$ in the top-antitop
propagators (\ref{propagator}) which is finite in the second
case ($\Gamma=\Gamma_t$), 
whereas in NWA one has $\Gamma\ar 0$. Correspondingly, in the 
integration over the phase space a Breit-Wigner distribution replaces
a delta  function. Moreover, in the case of the process \eettbbww\
there is an additional reduction (which increases with the top width)
due to the limited
phase space available, as the invariant mass tail
of the top-antitop decays at large $p^2$ in equation (\ref{propagator})  
can fall outside the kinematical bounds dictated by the CM energy 
and by the top mass \cite{tt}. Such a reduction gets larger
with increasing $m_t$ and decreasing $\MA$, whereas the effect is largely
independent of $\tan\beta$ (see Figs.~2a--b).
Differences are $\Ord13\%$, both at $\sqrt s=350$ and
400 GeV. The corrections due to the irreducible background
in \eebbww\ events can be estimated by comparing
rates in the second and third columns. These are larger as the top mass
and the CM energy of the collider increase, since, on the one hand,
non-resonant events do not
suffer from the threshold effects described above and, on the 
other hand, they do not have a pure $s$--channel structure like
top-antitop pair production. The maximum effects are $\approx5\%$
for $m_t=174$ GeV and $\approx9\%$ at $m_t=199$ GeV. In this case
there is not a marked dependence on $\tan\beta$.

By comparing the lower part with the upper part of Tabs.~I--II
one can appreciate the effect of the cut in $M_{\mathrm{had}}$, which
is indeed rather weak. This fact was already recognized in the case
of the \sm\ \cite{tt}, when the top mass was large.
Not even the increase here of the value adopted for $M_{\mathrm{had}}$
counterbalances this fact (in Ref.~\cite{Bagliesi} the
cut $M_{\mathrm{had}}>170$ GeV was used, for a top around 140 GeV).
The effect is however larger in the case of \eebbww, but only at
$\sqrt s=350$ GeV. 

The threshold scan in the region $2m_t-10~\mathrm{GeV}
\le\sqrt s-2~\mathrm{GeV}\le 2m_t+10~\mathrm{GeV}$ is performed 
in Fig.~3a--b, for $m_t=174$ and 199 GeV, respectively, for three
different values of $\MA$ and both \tba\ and 30.
From these figures is rather clear how the knowledge of 
\eebbww\ events that do not proceed through a $t\bar t$ resonance is 
essential is order to correctly estimate the sensitivity of the 
top excitation curve on $m_t$, $\Gamma_t$, $\alpha_s$, 
$M_\Phi$ and $\lambda_\Phi$ \cite{Bagliesi}. In fact, the difference
between the curves for the processes \eettbbww\ and \eebbww\ 
in Figs.~3a--b can be large below threshold.
However, the knowledge of the rates for the two above processes
allows one to perform a `subtraction'
between the dashed(dotted) and continuous(chain-dotted) curves
of Figs.~3a--b, thus extracting
the effect due to the irreducible background from the experimental samples
which will be used in the analyses.
We also notice that for $m_t=174$ GeV the cross sections below threshold 
are smaller than those 
for $m_t=199$ GeV, whereas above threshold 
it is the other way round: that is, the 
curves in Fig.~3a are steeper than those in Fig.~3b.  
Figs.~3c--d show the same rates when the ISR is included (we have used
the formulae given in Ref.~\cite{Nicro}).
Here, cross sections are smaller by several tens of percent than
those in Figs.~3a--b, with rates slightly more suppressed when
$\sqrt s\approx2m_t$. 
 
Other than on the integrated rates, finite width and irreducible background
effects are (strikingly) visible in the spectrum of the 
$W$ boson three-momentum, both at $\sqrt s=350$ and 400 GeV, and for
all top masses considered here.
Figs.~4a--b report the distributions in case of the NWA (at the bottom)
and of the complete set of diagrams in \eebbww\ events (two top plots),
for $\MA=100$ and 260 GeV and the following values of top masses:
$m_t=170,174$ GeV (Fig.~4a) and $m_t=195,199$ GeV (Fig.~4b).
As the shape of the distributions does not substantially 
change by varying $\tan\beta$,
for convenience we present here rates for the case \tba\ only.
The kinematic bounds on the $p_W$ distribution that would stick
out rather clearly in the case of the \eett\ process in NWA 
and would allow one to reconstruct $m_t$, are largely smeared out
once $\Gamma_t$ and background effects are properly included.
This is especially true for $m_t=174$ and 199 GeV. 
Thus, extreme care is needed in such analyses, together
with an accurate
knowledge of the above effects around the edges of the $p_W$ spectrum,
if one wants to achieve 
high precision measurements. 

The strong variations of the $p_W$ spectrum
with the top mass are also a reflection of its dependence on the fundamental
parameters of the \mssm, as can be seen from Figs.~5a--b 
(and again from Figs.~2a--b).
Here, the same distribution as in Figs.~4a--b is plotted, but for the
case of the complete \eebbww\ process, and for different values 
of $\MA$ 
(with $m_t$ fixed at 174 and 199 GeV, in Fig.~5a and 5b, respectively).
For the same reasons as above, we plot here distributions for only one
of the two values of $\tan\beta$ that we have chosen (for example, \tbb\ now).
The same sort of trend would also be visible in the cases $m_t=170,172$ and
195, 197 GeV. 
The normalisation of the various curves is rather different, 
according to the numbers given in Tabs.~I--II
and depending on the
value of $\MA$, when the latter is less than $\approx180$ GeV (compare
to Figs.~4a--b too). For $\MA\OOrd180$ GeV there is a clear `degeneracy'
between the different curves, reflecting the fact, on the one hand,  that
the value of $\Gamma_t$ does not change in this interval and, on the other
hand, that the differences due to the different contributions of  
the Higgs processes to the total cross section are rather small (compared
to $e^+e^-\ar t\bar t$). It is however clear that any possible deviation from
the shapes and normalizations of the cross sections predicted 
by the \mssm\ (at fixed $\MA$ and $\tb$) can easily be tested by studying
the $p_W$ spectrum. 

As explained in the Introduction, 
top studies at the NLC will also be performed far above threshold.
Thus, we report in Tab.~III the cross sections for processes
1.--3.  at $\sqrt s=500$ GeV, assuming for the top mass the two
values $m_t=174$ and 199 GeV (in this energy regime there is practically
no dependence on $m_t$ if this changes by 2 or 4 GeV). The selection
of Higgs masses in Tab.~III is the same as in the two previous ones, as well
as the organisation of the rates depending on $\MA$, $\tb$ and on the cuts. 
Since the suppression of the high invariant masses of
the $bW$ pairs does not act any longer, finite width effects
above threshold are fairly small, of a few percent only.
In some instances these are even completely negligible: for example,
for $m_t=174$ GeV. Irreducible background effects are around $5\%$
at the most, generally for all combinations of $m_t$, $\MA$ and $\tb$.
The cut implemented here is $0.95\le x_E\le1.05$,
where $x_E=E_{bW\ar3\mathrm{jet}}/E_{\mathrm{beam}}$, 
again according to the analysis of Ref.~\cite{Bagliesi}.
This constraint is somewhat more effective that the
one used at threshold, as it reduces the \eebbww\ rates 
 by more than twice the \eettbbww\ ones (by approximately $5\%$ the
former and by $2\%$ the latter).

Above threshold one can extract the value of the top mass
from a fit to the observed invariant mass distributions of the 
three-jet systems $bW^+$ (the `right' one) and $\bar bW^+$ (the `wrong'
one). Figs.~6a--b show such spectra for $\sqrt s=500$ GeV,
$m_t=174$ and 199 GeV, $\MA=100$ and 260 GeV, and for \tba\ (Fig.~6a)
and \tbb\ (Fig.~6b). As can be seen clearly, the $\bar bW^+$ distribution
is generally well below the $bW^+$ one around $m_t$, by more than one order
of magnitude. One should notice however that the bins in Figs.~6a--b are
2 GeV wide, when the experimental resolutions (in energy and 
angle) on the hadronic system constituted   by the three jets will
probably be worse than that, such that Figs.~6a--b represent a sort of
`benchmark' result of the data analyses.
For example, an experimental resolution in $M_{\mathrm{3jet}}$ of only (let us 
say) 10 GeV would correspond to lowering the $bW^+$ peaks in Figs.~6a--b
by a factor of five, leaving the spectrum in $\bar bW^+$ unchanged.

\subsection*{3.2 The processes $e^+e^-\ar ZH\ar (b\bar b)(W^+W^-)$,
$e^+e^-\ar AH\ar (b\bar b)(W^+W^-)$ and $e^+e^-\ar hW^+W^-\ar (b\bar b)W^+W^-$}

The neutral Higgs bosons of the \mssm\ can be produced primarily at the NLC 
via
\be\label{Higgs1}
\mathrm{bremsstrahlung:} \qquad e^+e^-\ar (Z^*)\ar Z\Phi; 
\ee
\be\label{Higgs2}
\mathrm{pair~production:} \qquad e^+e^-\ar (Z^*)\ar A\Phi; 
\ee
\be
WW\mathrm{-fusion:} \qquad e^+e^-\ar \nu_e\bar\nu_e(W^*W^*)
                      \ar \nu_e\bar\nu_e\Phi;  
\ee
\be
ZZ\mathrm{-fusion:} \qquad e^+e^-\ar e^+e^-(Z^*Z^*)\ar e^+e^-\Phi;  
\ee
where here $\Phi=h$ and $H$\footnote{The pseudoscalar boson $A$
does not couple at leading order to the gauge bosons $W$ and $Z$.}. 
Simple expressions
relate their cross sections to those of the bremsstrahlung and vector-vector
fusion processes of the \sm\ \cite{expre}.
Updated and detailed
theoretical \cite{DKZ,BCDKZ,Orange3} (also at the complete one-loop 
level \cite{Poland})
and experimental \cite{Janot} analyses making use
of the above processes are available in the literature. 
In general, the first two mechanisms dominate over the last two
at smaller CM energies ($\sqrt s \approx300-500$ GeV), whereas at
$\sqrt s\OOrd 500$ GeV $WW$-fusion has bigger rates if the $\Phi$ mass 
is less than $\approx160$ GeV. In general, the cross section for 
$ZZ$-fusion is about one order of magnitude smaller than that for 
$WW$-fusion. 
Furthermore, from the results of Ref.~\cite{WWh}, one deduces that 
sizeable rates should be expected for events of 
\be\label{Higgs3}
\mathrm{Higgs~production~with~two}~{W}~\mathrm{bosons:} 
\qquad e^+e^-\ar hW^+W^-,
\ee
especially at small value of $\Mh$ and $\tb$ \cite{guide}.

By studying \eebbww\ events one implicitly considers the
first two processes (\ref{Higgs1})--(\ref{Higgs2}) and the last one 
(\ref{Higgs3}), in which $h,A,Z\ar b\bar b$ and $H\ar 
W^+W^-$\footnote{Also other Higgs production mechanisms take place
in the whole of the process \eebbww\ (see graphs 20--21 \& 27 in
Refs.~\cite{tt,ZH}): these are however very much suppressed.}. 
As we are concerned here with final-state $W$ bosons produced 
on-shell, we are forced to ignore the off-shell decays
$h,H\ar W^{*}W^{*}$ (that is, when $M_{h,H}<2M_W$).
However, the below-threshold channel $W^*W^*$ is unlikely to be 
particularly relevant in experimental analyses. 
In the case of the light neutral Higgs the corresponding BR is either too small
(at small $\tan\beta$) or is sizable only over a very narrow portion
of the \mssm\ parameter space (at large $\tan\beta$).
In the case of the heavy neutral Higgs, the decay 
$H\ar W^*W^*$ is overwhelmed by $H\ar hh$ 
below the $2M_W$ threshold for $\tan\beta=1.5$,
whereas for $\tan\beta=30.$ it is always smaller than $H\ar hh,AA$ as well as
$H\ar b\bar b,\tau^+\tau^-$, in the relevant off-shell range.
The on-shell decay $H\ar WW$ is certainly 
of interest for small values of $\tan\beta$, as it dominates
over the range $160~\mbox{GeV}\Ord M_H\Ord 200~\mbox{GeV}$ and has 
the second highest BR after the $hh$ channel 
from $\MA\approx200$ GeV up to the opening
of the threshold for the decay $H\ar t\bar t$.
We do not expect any sizable $H\ar WW$ signal if $\tan\beta$ is 
on the contrary very large \cite{ioejames}. 

Concerning the $b\bar b$ decays of the $Z$, $h$ 
and $A$ bosons, we stress two aspects. On 
the one hand, the same arguments which led us in the Introduction
to count the $Z\ar b\bar b$
decay mode among the cleanest $Z$ signatures in the \sm\ at the NLC
are certainly still valid within the \mssm.
On the other hand, the 
$h,A\ar b\bar b$ decay channels largely dominate the decay spectrum
of the two Higgs scalars: in fact, the $BR(h\ar b\bar b)$ is the largest
for all values of $\MA$ and $\tan\beta$ whereas the $BR(A\ar b\bar b)$ 
is overcome by that of other channels only at small
$\tan\beta$'s and for $\MA\OOrd200$ GeV \cite{ioejames}.
Moreover, we would like to stress that in the $W^+W^-$ semi-leptonic decay mode
we do not expect complications due the jet combinatorics, because,
first,
excellent $b$-tagging performances are expected at 
the NLC \cite{ideal} and, second, $b$-decays of the $W$ bosons
are prohibited by the top mass $m_t>M_W$ and the
Cabibbo-Kobayashi-Maskawa mixing matrix.

Among the various search strategies that can be adopted
at the NLC in order to detect such Higgs signals \cite{ee500} 
we consider here the following two:
\begin{itemize}
\item the `missing mass' analysis; 
\item the `direct reconstruction' method.
\end{itemize}
They are both described in Ref.~\cite{grosse}, in the case of the \sm; however,
these procedures are useful also in the case of the \mssm.
In this context, 
in the first case one studies the invariant mass recoiling against
the $Z$ and $A$ bosons tagged via the $b\bar b$ pair, by means of the
relation $M_{\mathrm{recoil}}^2=[(p_{e^+}+p_{e^-})-(p_{b}+p_{\bar b})]^2$,
whereas in the second case one reconstructs Higgs peaks  
by computing the resonant invariant mass directly 
from the four-momenta of the decay products of the scalar 
bosons \cite{BCDKZ,Orange3}\footnote{In our 
`partonic' analysis (where no detector effect and
experimental efficiencies are considered) 
the two spectra in 
$M_{\mathrm{recoil}}$ and $M_{WW}$ necessarily coincide.}.
Thus, both procedures are applicable to processes 
(\ref{Higgs1})--(\ref{Higgs2}), whereas in case of process (\ref{Higgs3})
only the second is exploitable. 

Our results concerning Higgs physics in \eebbww\ events are 
summarised in Figs.~7--11. As already announced, Higgs rates
are expected to be larger at smaller \tb\ values, for all the
processes (\ref{Higgs1})--(\ref{Higgs2}) and (\ref{Higgs3}). 
Therefore, we will treat
the case \tba\ only. 

As explained in Ref.~\cite{ZH},  looking at the
plane $(M_{b\bar b},M_{W^+W^-})$ one realises that,
whereas Higgs signals 
tend to concentrate in single
clusters whose size is determined in the end by the experimental
reconstruction uncertainties,
on the contrary,
background events (mostly from $t\bar t$ production and decay) tend
to fill a rather large kinematical region. 
In Fig.~7 we show the 
boundaries of the double differential distribution
$d\sigma/dM_{b\bar b}/dM_{W^+W^-}$ in the plane
$(M_{b\bar b},M_{W^+W^-})$ in the case of
$e^+e^-\rightarrow t\bar t$ in NWA, for all the combinations of $\sqrt s$
and $m_t$ that we have adopted in our study, without any sort of
cuts\footnote{The
dependence of such regions on the actual value of $\Gamma_t$ (thus
on the \mssm\ parameters) in
eq.~(\ref{propagator}), when $\Gamma\ar0$, is completely 
negligible. Rates in Fig.~7
have been plotted for $\MA=100$ GeV and \tba.}.

As outlined in Refs.~\cite{ZH}, it is then possible,  
for a given $m_t$, to find
values of the \mssm\ neutral Higgs masses for which the two
double differential distributions (of Higgs and top events)
do not overlap. In these cases the latter 
can be very simply distinguished from
the former and detecting the Higgs scalars is only a matter
of event rate, whereas when the $M_{b\bar b}$ and/or $M_{W^+W^-}$
spectra overlap, more refined studies are necessary.
Furthermore, from Fig.~7 it also clear that it should
be easier in principle to detect Higgs signals when the CM energy of
the collider is set around the $2m_t$ threshold, more than when the former
is far beyond the latter. In fact, first, rates for \eett\ are smaller at
threshold that at $\sqrt s=500$ GeV and, second, the regions in the
$(M_{b\bar b},M_{W^+W^-})$ plane occupied
by \ttb\ production and decay are more narrow (at fixed $m_t$).
Therefore, to some extent, a collider energy configuration primarily
designed for top studies at threshold (i.e., $\Delta E\ar0$) 
would also improve the chances of successful Higgs searches 
in the $b\bar bW^+W^-$ channel.

As an example of the various possible cases we study here the combinations
($\tba$):
\begin{enumerate}
\item $\sqrt s=350$ GeV, $m_t=174$ GeV and $\MA=140,180$ GeV;
\item $\sqrt s=400$ GeV, $m_t=199$ GeV and $\MA=140,180$ GeV;
\item $\sqrt s=400$ GeV, $m_t=199$ GeV and $\MA=60,100,220,260,300,340$ GeV;
\item $\sqrt s=500$ GeV, $m_t=199$ GeV and $\MA=60,100$ GeV.
\end{enumerate}

In cases 1. and 2. the Higgs peaks are right next to the maxima of
the $t\bar t$ distributions in both the $M_{b\bar b}$ and $M_{W^+W^-}$
spectra (Figs.~8--9). For the first combination the following Higgs peaks
are expected:
$h\ar b\bar b$ at $M_{b\bar b}\approx80(93)$ and 86(102) GeV, 
corresponding to 
$m_t=174(199)$ GeV; $A\ar b\bar b$ at $M_{b\bar b}\approx140$ and 180
GeV; $H\ar W^+W^-$ at $M_{W^+W^-}\approx182(195)$ and 212(222) GeV, again 
corresponding to $m_t=174(199)$ GeV.
The $Z\ar b\bar b$ resonance gives indirect evidence of process
(\ref{Higgs1}) with $\Phi=h$, yet many of the non-Higgs diagrams proceed
through a $Z$ decay into $b\bar b$ pairs, such that the two normalisations
(of signal and background events) need to be carefully known in order
to make Higgs detection feasible in such a case.
The combination of CM energy and top/Higgs masses
where the rewards for Higgs searches are largest is the second.
In this case Higgs peaks in the $b\bar b$ spectrum
are already visible before any selection cut
is applied, apart from the cases: $M_h\approx93$ GeV (which is degenerate
with the $Z$ peak) and $M_A\approx180$ GeV (overwhelmed by the
$t\bar t$ background), Fig.~8b (main plot). On the contrary, 
for the combination in 1. it is generally impossible
to disentangle Higgs resonances from the pure total cross sections before
cuts, Fig.~8a (main plot).
These conclusions depend largely on the 
invariant mass resolutions, however, it has to be observed that 
bins in the main windows of Fig.~8--9 are large, 5 GeV (a value, in our
opinion, which is not too distant from the 
performances experimentally achievable).
Of course, if the precision in the angular and energy measurements is
higher (resulting, let us say, in a mass resolution 
around 1 GeV), all the above decay phenomenology
in $b\bar b$ events can be covered in principle (small windows
in Figs.~8a--b), with special attention
devoted to the case of $h/Z$ mass degeneracy. 
In the case of the $H$ resonance in the $M_{W^+W^-}$ spectrum things
are less optimistic (Figs.~9a--b, main windows), apart from the
case $M_{H}\approx195$ GeV (at $\sqrt s=400$ GeV, for $\MA=140$ GeV).
Therefore, 
in order to disentangle $H\ar W^+W^-$ resonances, one generally needs
to apply Higgs selection cuts. For example, by requiring that in events
for which $|M_{b\bar b}-M_{Z,A}|<10$ GeV one of the $W$'s fails
to reproduce the kinematics
of a \ttb\ final state when coupled with either of the two $b$'s (namely that
$m_t-10~{\mathrm{GeV}}>|M_{W^+b(W^+\bar b)}|>m_t+10~{\mathrm{GeV}}$
and $E_{\mathrm{beam}}-10~{\mathrm{GeV}}>|E_{W^+}+E_{b(\bar b)}|>
E_{\mathrm{beam}}+10~{\mathrm{GeV}}$),
one obtains the results displayed in the two small insertions on the right
of Figs.~9a--b. That is, to recognise $H$ signals in $W^+W^-$ decays
one needs both to apply selection cuts to reduce the top-antitop background
and to achieve high invariant mass resolutions to resolve the peaks (unless
$M_{H}$ is around $195$ GeV when  $\MA=140$ GeV and the CM energy
is 400 GeV).
Furthermore, it is clear that once similar cuts are applied to the 
$M_{b\bar b}$ spectrum, all the decay phenomenology for
$h,A\ar b\bar b$ should
be exploited easily.

From the rates given in Figs.~8 
one can generally extract about ten Higgs events per year with
rather little background. 
On the contrary, from the numbers in Figs.~9 (when $M_H\not\approx195$ GeV),
one is able to detect Higgs signals only after several years of running
and if the collider yearly luminosity is of the order of 
$100~\mathrm{fb}^{-1}$. It should also be noticed that in the case of the 
$H\ar W^+W^-$ decay the application of Higgs selection cuts reduces even 
further the total rates.
Finally , one has to ultimately consider that
the semi-leptonic signature $W^+W^-\ar(\mathrm{jj})
(\ell\bar\nu_\ell)$, with $\ell=e,\mu$, introduces a reduction factor
equal to the product of the $BR(W\ar\mathrm{jj})\approx70\%$ and
the $BR(W\ar\ell\bar\nu_\ell)\approx20\%$: actually, times 2, because 
of the two combinations in which the $W$'s can decay 
semi-leptonically. 

The combination in 3. corresponds to the case in which the $h,A\ar b\bar b$
peaks are either on the tail of the top-antitop distribution in $M_{b\bar b}$
(Fig.~10, upper window) or rather far from the $Z$ resonance
such that problems of mass degeneracy should be avoided (Fig.~10, two 
lower windows). Chances of $h$ and $A$ detection are 
convincing (more at smaller than at larger values of $\MA$),
provided that good invariant mass resolutions can be achieved (distributions
in Fig.~10 are plotted in bins of 1 GeV).
For these combinations of CM energy and top masses 
$H\ar W^+W^-$ signals can be disentangled only if $\MA\Ord100$ GeV,
for which $M_H$ is close enough to the $2M_W$ threshold, such that 
the BR into two $W$'s is large (the pattern is rather similar to that
illustrated for $\MA=140$ GeV). For $\MA\OOrd220$ GeV the $H$ boson
is very heavy 
(i.e., $\MH\OOrd250$ GeV) compared to the available CM energy
at threshold, and so the corresponding cross section is kinematically
suppressed. 

Fig.~11 shows what happens to the $M_{W^+W^-}$ spectrum
for the combination of masses 
in 4., with the collider energy well above the $t\bar t$ 
threshold (contrary to the previous three cases). Here, even when
no selection cut is applied, $H$ signals are visible, again especially at small
values of $\MA$ (i.e., for $\MH$ rather
close to the $2M_W$ threshold). The $H\ar W^+W^-$ peaks, in fact,
occur in Fig.~11 at $M_{W^+W^-}\approx165$ and 176 GeV.
In this case then, the chances of heavy neutral Higgs detection are
larger, thanks especially to the fact that the region delimited by
$60~\mathrm{GeV}\Ord M_{b\bar b}\Ord100~\mathrm{GeV}$ 
and $M_{W^+W^-}$ between the above two values is practically free
from the background due to \eett\ events (see Fig.~7).

\section*{4. Conclusions} 

In this paper we have studied the process
$$
e^+e^-\ar b\bar bW^+W^-
$$
within the \mssm, at NLC energies ($\sqrt s=350,400$ and 500 GeV),
for the following values of the parameters of the model:
$60~\mathrm{GeV}\Ord\MA\Ord340~\mathrm{GeV}$ and $\tb=1.5, 30$.
In our calculation, all the Feynman diagrams contributing at tree-level  
to the above reaction have been considered, with no approximations.

This process is phenomenologically extremely interesting, as it includes
among the various contributions to the total cross section the following
subprocesses:
$$
e^+e^-\ar t\bar t\ar (bW^+) (\bar bW^-),
$$
$$
e^+e^-\ar ZH\ar (b\bar b) (W^+W^-), 
$$
$$
e^+e^-\ar AH\ar (b\bar b) (W^+W^-), 
$$
$$
e^+e^-\ar hW^+W^- \ar (b\bar b) W^+W^-, 
$$
where $H,h$ and $A$ represent the neutral Higgs bosons of the \mssm, 
as well the contributions of non-resonant diagrams (irreducible background).
For the top mass, the following values have been adopted:
$m_t=170,172$ and 174 GeV (when $\sqrt s=350$ GeV) and
$m_t=195,197$ and 199 GeV (when $\sqrt s=400$ GeV).
At $\sqrt s=500$ GeV the two values $m_t=174$ and 199 GeV have 
been considered.
Our results are applicable to all energies excluding a narrow
window of a few GeV at threshold.

Final states of the type $b\bar bW^+W^-$ are decisive in both
top and Higgs processes. On the one hand, top pairs produced
via the process \eett\ decay 
through $t\bar t\ar (bW^+)(\bar bW^-)$ over most 
of the \mssm\ parameter space $(\MA,\tb)$. On the
other hand, the $b\bar b$ decay channel might well be one of best
ways to detect the $Z$ gauge boson and represents the most 
accessible signature 
of the Higgs scalars $h$ and $A$ (thanks to the corresponding
high BRs and to the excellent performance in tagging jets originating 
from $b$-quarks now foreseen for the vertex devices which 
will be installed at the NLC). Moreover, the decay channel $H\ar W^+W^-$
of the heaviest of the \mssm\ Higgs scalars is dominant over
a sizable part of the $\MH$ range, for small $\tb$'s.  

Phenomenological analyses have been carried out, in order to quantify, first,
the influence of finite top width and irreducible
background effects on the integrated and differential rates as obtained from 
\eett\ events in Narrow Width  Approximation and, second,
to establish the detectability of the Higgs processes, as a function 
of the 
values assumed by the fundamental parameters of the \mssm.

Although 
before drawing any firm conclusion from our results, these should be folded
with a realistic simulation of the expected performances of the NLC detectors,
they clearly indicate that the finite width of the top and the irreducible
background can have a significant impact on the measurement of the parameters
of the top, both near and above the $\sqrt s=2m_t$ threshold. 
Furthermore, in a number of cases (especially at small $\MA$'s and $\tb$'s),
it should be possible to disentangle Higgs resonances, 
the $h,A\ar b\bar b$ channels being
easier to detect than the $H\ar W^+W^-$ one.

Throughout our analysis we have considered the $W^+W^-\ar (\mathrm{jj})
(\ell\bar
\nu_\ell)$  signature of the gauge bosons (with $\ell=e,\mu$), and assumed
high $b$- and lepton tagging efficiency, such that all the background to
the mentioned top and Higgs events can be ascribed to the irreducible one.

The effect of Initial State Radiation has been analysed in a few cases.

Finally, a similar study of the pure \mssm\ processes
$$
e^+e^-\ar b\bar b W^+H^- 
$$
and
$$
e^+e^-\ar b\bar b H^+H^-
$$
is now in progress \cite{preparation}.

\section*{5. Acknowledgements}

We thank Bryan Webber for reading the manuscript.
This work is supported in part by the
Ministero dell' Universit\`a e della Ricerca Scientifica, the UK PPARC,
and   the EC Programme
``Human Capital and Mobility'', Network ``Physics at High Energy
Colliders'', contract CHRX-CT93-0357 (DG 12 COMA).

\goodbreak

\vfill
\newpage

\subsection*{Table Captions}
\begin{description}
\item[{\bf Tab.~I}  ] Cross sections in femtobarns
for $e^+e^-\ar t\bar t$ (Narrow
Width Approximation, $\Gamma\ar0$), for
$e^+e^-\ar t\bar t\ar b\bar b W^+W^-$ (production and decay diagram
only, $\Gamma=\Gamma_t$) and for \eebbww\ (all diagrams at tree-level,
$\Gamma=\Gamma_t$), within the \mssm,
at $\sqrt s=350$ GeV, for $m_t=170,172$ and 174 GeV, 
and $\MA=60(140)[220]\{300\}$ GeV. The upper part refers to rates obtained
when no cut is applied, the lower one to rates obtained when the
cut $M_{\mathrm{had}}>200$ GeV is implemented. The first(second)
row reports rates for \tba(30.).

\item[{\bf Tab.~II} ] Same as in Tab.~I, 
at $\sqrt s=400$ GeV, for $m_t=195,197$ and 199 GeV.
 
\item[{\bf Tab.~III}] Cross sections in femtobarns
for $e^+e^-\ar t\bar t$ (Narrow
Width Approximation, $\Gamma\ar0$), for
$e^+e^-\ar t\bar t\ar b\bar b W^+W^-$ (production and decay diagram
only, $\Gamma=\Gamma_t$) and for \eebbww\ (all diagrams at tree-level,
$\Gamma=\Gamma_t$), within the \mssm,
at $\sqrt s=500$ GeV, for $m_t=174$ and 199 GeV, 
and $\MA=60(140)[220]\{300\}$ GeV. The upper part refers to rates obtained
when no cut is applied, the lower one to rates obtained when the
cut $0.95\le x_E\le1.05$ is implemented. The first(second)
row reports rates for \tba(30.).
 
\end{description}

\vfill
\newpage

\subsection*{Figure Captions}
\begin{description}
\item[{\bf Fig.~1} ] The \mssm\ Feynman graph contributing in lowest order
to $e^+e^-\ar (Ah),(AH)\ar b\bar b W^+W^-$.
The dashed line
connected to the $b\bar b$ one represents the $A$ scalar, whereas 
the other refers to both a $h$ and a $H$ boson. The other 27 
\sm-like topologies
of Feynman graphs contributing in lowest order to the complete
ME for \eebbww\ in the
\mssm\ can be found in Refs.~\cite{tt,ZH}, where a dashed line
represents both a $h$ and a $H$ boson.

\item[{\bf Fig.~2} ] \mssm\ mass relations of the charged $H^\pm$
and of the neutral $h$ and $H$ Higgs bosons with respect to
the pseudoscalar neutral one $A$,
for $\tan\beta=1.5$ and 30., for different values of the top mass ({\bf a}),
and partial decay widths of the top quark in the channels $t\ar bH^\pm$
and $t\ar bW^\pm$ within the \mssm, as a function of $\MA$, for 
$\tan\beta=1.5$ and 30., and different values of top masses  ({\bf b}).
Continuous and dotted lines in {\bf a} coincide, as we have used here the
tree-level relation for the charged Higgs mass. 
The dashed and chain-dotted curves corresponding to the $h(H)$ boson
are those extending to the right lower(upper) corner in {\bf a}. 

\item[{\bf Fig.~3} ] Cross section in femtobarns for \eebbww\ events
around the top-antitop threshold, for two
different values of $\MA$: 
({\bf a}) $\sqrt s=350$ GeV and $m_t=174$ GeV; 
({\bf b}) $\sqrt s=400$ GeV and $m_t=199$ GeV; 
Continuous lines: \tba\ and 
production and decay diagram only. 
Dashed lines: \tba\ and
all diagrams. 
Dotted lines: \tbb\ and 
production and decay diagram only. 
Chain-dotted lines: \tbb\ and
all diagrams. In all cases $\Gamma=\Gamma_t$ has been used. 
In ({\bf c}) and ({\bf d}) the same as above, but in the presence of ISR.
Continuous and dotted lines coincide for $\MA\OOrd180$ GeV.

\item[{\bf Fig.~4} ] Differential distributions in the momentum of
the $W$ boson, in \eebbww\ events, within the \mssm, 
in the cases: Narrow Width Approximation for $\MA=100$ and
260 GeV (lower windows), all diagrams for $\MA=100$ GeV (upper windows)
and $\MA=260$ GeV (middle windows), with $\tan\beta=1.5$.
The CM energies and the top masses are:
({\bf a}) $\sqrt s=350$ GeV and $m_t=170,174$ GeV;
({\bf b}) $\sqrt s=400$ GeV and $m_t=195,199$ GeV.
Continuous and dotted lines: $m_t=170(195)$ GeV.
Dashed and chain-dotted lines: $m_t=174(199)$ GeV.
The cut in hadronic mass $M_{\mathrm{had}}>200$ GeV is negligible 
in all plots.

\item[{\bf Fig.~5} ] Differential distributions in the momentum of
the $W$ boson, in \eebbww\ events, within the \mssm, for a selection
of pseudoscalar Higgs masses, with $\tan\beta=30.$
All diagrams are here
considered. The CM mass energies and top masses are:
({\bf a}) $\sqrt s=350$ GeV and $m_t=174$ GeV;
({\bf b}) $\sqrt s=400$ GeV and $m_t=199$ GeV.
No cut has been applied.

\item[{\bf Fig.~6} ] Differential distributions in the invariant
mass of the {`right'} and {`wrong'} three-jet combinations (see the text),
in \eebbww\ events, within the \mssm, for $\MA=100$ (upper window)
and 260 (lower window) GeV, 
with $\tan\beta=1.5$ ({\bf a}) and $30.$ ({\bf b}). All diagrams have been here
considered. The CM energy is 
$\sqrt s=500$ GeV. 
Continuous lines: $m_t=174$ GeV.
Dotted lines: $m_t=199$ GeV.
No cut has been applied.

\item[{\bf Fig.~7} ] The boundaries of the double differential distributions
$d\sigma/dM_{b\bar b}/dM_{W^+W^-}$ in the plane
$(M_{b\bar b},M_{W^+W^-})$ for $e^+e^-\rightarrow t\bar t$ events
in NWA,
at $\sqrt s=350, 400$ and $500$ GeV, 
for different values of $m_t$.
No cut has been applied here.
 
\item[{\bf Fig.~8} ] Differential distributions in invariant mass
of the $b\bar b$ pair, in \eebbww\ events, within the \mssm,
for $\tan\beta=1.5$ GeV and two different
values of $\MA$. CM energies and top masses are:
$\sqrt s=350$ GeV and $m_t=174$ GeV ({\bf a});
$\sqrt s=400$ GeV and $m_t=199$ GeV ({\bf b}).
Continuous lines: all \eebbww\ diagrams, with $\MA=140$ GeV.  
Dashed lines: all \eebbww\ diagrams, with $\MA=180$ GeV.  
The two small windows on the right 
show the regions around the $h$ and $Z$ resonances 
enlarged and plotted with high resolution, together 
with the $t\bar t$ contribution in NWA (shaded). 
The small central window in {\bf b} does the same for the $A$ resonance, at 
$M_{b\bar b}=140$ GeV.
No cut has been applied here.

\item[{\bf Fig.~9} ] Differential distributions in invariant mass
of the $W^+W^-$ pair, in \eebbww\ events, within the \mssm,
for $\tan\beta=1.5$ GeV and two different
values of $\MA$. CM energies and top masses are:
$\sqrt s=350$ GeV and $m_t=174$ GeV ({\bf a});
$\sqrt s=400$ GeV and $m_t=199$ GeV ({\bf b}).
In the two small windows the regions around the $H$ resonances are
enlarged and plotted with high resolution, together 
with the $t\bar t$ contribution in NWA (shaded).
Continuous lines: all \eebbww\ diagrams, with $\MA=140$ GeV.   
Dashed lines: all \eebbww\ diagrams, with $\MA=180$ GeV.  
In the large window no cut is applied, whereas in the two small ones the
following constraints have been implemented:
$m_t-10~{\mathrm{GeV}}>|M_{W^+b(W^+\bar b)}|>m_t+10~{\mathrm{GeV}}$
and $E_{\mathrm{beam}}-10~{\mathrm{GeV}}>|E_{W^+}+E_{b(\bar b)}|>
E_{\mathrm{beam}}+10~{\mathrm{GeV}}$, $|M_{b\bar b}-M_{Z,A}|<10$ GeV.

\item[{\bf Fig.~10}] Differential distributions in invariant mass
of the $b\bar b$ pair, in \eebbww\ events, within the \mssm,
for $\tan\beta=1.5$ GeV and six different
values of $\MA$. CM energies and top masses are:
$\sqrt s=400$ GeV and $m_t=199$ GeV.
The shaded areas correspond to the $t\bar t$ contribution in NWA:
in the top window the dotted(black) region refers to 
$\MA=100(60)$ GeV, whereas in the central(lower) window the shaded
regions refer to   
$\MA=220,260(300,340)$ GeV.
All diagrams have been considered in the resonant contributions.
No cut has been applied here.

\item[{\bf Fig.~11}] Differential distributions in invariant mass
of the $W^+W^-$ pair, in \eebbww\ events, within the \mssm,
for $\tan\beta=1.5$ GeV and two different
values of $\MA$. CM energies and top masses are:
$\sqrt s=500$ GeV and $m_t=199$ GeV.
The shaded areas correspond to the $t\bar t$ contribution in NWA.
All diagrams have been considered in the resonant contributions.
No cut has been applied here.

\end{description}

\vfill
\clearpage

\begin{table}
\begin{center}
\begin{tabular}{|c|c|c|c|}
\hline
\rule[0cm]{0cm}{0cm}
$m_t$ (GeV)                           &\omit  
$~$                                   &\omit  
$\sigma(e^+e^-\ar X)$ (fb)            &
$~$                                   \\ \hline  \hline
\rule[0cm]{0cm}{0cm}
$~$                                  &  
$t\bar t$                            &  
$t\bar t \rightarrow \bar b bW^+W^-$ & 
$\bar b bW^+W^-$                     \\ \hline\hline
\rule[0cm]{0cm}{0cm}
$170$ &  
$341(503)[508]\{508\}$  &  
$314(470)[475]\{475\}$  &    
$322(481)[484]\{484\}$  \\
                             &  
$311(502)[508]\{508\}$  &  
$286(469)[475]\{475\}$  &  
$292(479)[484]\{484\}$  \\ \hline
\rule[0cm]{0cm}{0cm}
$172$ &  
$264(390)[396]\{396\}$  &  
$237(357)[362]\{362\}$  &  
$244(366)[371]\{374\}$  \\
                             &  
$243(389)[396]\{396\}$  &  
$217(355)[362]\{362\}$  &  
$224(363)[370]\{370\}$  \\ \hline
\rule[0cm]{0cm}{0cm}
$174$ &  
$152(225)[230]\{230\}$  &  
$134(200)[204]\{204\}$  &  
$141(210)[213]\{214\}$  \\
                             &  
$141(224)[230]\{230\}$  &  
$124(199)[204]\{204\}$  &  
$130(207)[212]\{212\}$  \\ \hline
\multicolumn{4}{|c|}
{\rule[0cm]{0cm}{0cm}
no cut}
 \\ \hline  \hline
\rule[0cm]{0cm}{0cm}
$170$ &  
$341(503)[508]\{508\}$ &  
$312(467)[472]\{472\}$ &  
$318(476)[479]\{479\}$ \\
                             &  
$311(502)[508]\{508\}$  &  
$283(466)[472]\{472\}$  &  
$289(474)[479]\{479\}$  \\ \hline
\rule[0cm]{0cm}{0cm}
$172$ &  
$264(390)[396]\{396\}$  &  
$235(355)[360]\{360\}$  &  
$241(362)[367]\{370\}$ \\
                             &  
$243(389)[396]\{396\}$  &  
$215(353)[360]\{360\}$  &  
$221(360)[366]\{366\}$  \\ \hline
\rule[0cm]{0cm}{0cm}
$174$ &  
$152(225)[230]\{230\}$  &  
$132(199)[203]\{203\}$  &  
$136(207)[210]\{210\}$ \\
                             &  
$141(224)[230]\{230\}$  &  
$123(198)[203]\{203\}$  &  
$127(204)[209]\{209\}$  \\ \hline
\multicolumn{4}{|c|}
{\rule[0cm]{0cm}{0cm}
$M_{{\mathrm{had}}}>200$ GeV}
 \\ \hline  \hline
\multicolumn{4}{|c|}
{\rule[0cm]{0cm}{0cm}
$\sqrt s=350$ GeV}
 \\ \hline 
\multicolumn{4}{c}
{\rule{0cm}{1cm}
{\Large Tab. I}}  \\
\multicolumn{4}{c}
{\rule{0cm}{0cm}}
\end{tabular}
\end{center}
\end{table}

\vfill
\clearpage

\begin{table}
\begin{center}
\begin{tabular}{|c|c|c|c|}
\hline
\rule[0cm]{0cm}{0cm}
$m_t$ (GeV)                           &\omit  
$~$                                   &\omit  
$\sigma(e^+e^-\ar X)$ (fb)            &
$~$                                   \\ \hline  \hline
\rule[0cm]{0cm}{0cm}
$~$                                   &  
$t\bar t$                             &  
$t\bar t \rightarrow \bar b bW^+W^-$  &  
$\bar b bW^+W^-$                      \\ \hline\hline
\rule[0cm]{0cm}{0cm}
$195$                        &  
$227(329)[362]\{362\}$  &  
$201(298)[329]\{329\}$  &  
$214(311)[340]\{340\}$  \\
                             &  
$231(331)[362]\{362\}$  &  
$204(299)[329]\{329\}$  &  
$213(311)[340]\{340\}$  \\ \hline
\rule[0cm]{0cm}{0cm}
$197$ &  
$176(254)[282]\{282\}$  &  
$151(224)[249]\{249\}$  &  
$164(235)[261]\{261\}$  \\
                             &  
$180(256)[282]\{282\}$  &  
$155(225)[249]\{249\}$  &  
$164(238)[260]\{261\}$  \\ \hline
\rule[0cm]{0cm}{0cm}
$199$ &  
$102(147)[164]\{164\}$  &  
$89(129)[145]\{145\}$  &  
$101(144)[156]\{155\}$  \\
                             &  
$105(148)[164]\{164\}$  &  
$92(130)[145]\{145\}$  &  
$101(142)[155]\{156\}$  \\ \hline
\multicolumn{4}{|c|}
{\rule[0cm]{0cm}{0cm}
no cut}
 \\ \hline  \hline
\rule[0cm]{0cm}{0cm}
$195$ &  
$227(329)[362]\{362\}$ &  
$200(297)[328]\{328\}$ &  
$213(310)[339]\{339\}$ \\
                             &  
$231(331)[362]\{362\}$  &  
$204(299)[329]\{329\}$  &  
$212(310)[339]\{339\}$  \\ \hline
\rule[0cm]{0cm}{0cm}
$197$ &  
$176(254)[282]\{282\}$ &  
$150(223)[249]\{249\}$  &  
$163(234)[260]\{260\}$ \\
                             &  
$180(256)[282]\{282\}$  &  
$154(224)[249]\{249\}$ &  
$163(237)[259]\{260\}$  \\ \hline
\rule[0cm]{0cm}{0cm}
$199$ &  
$102(147)[164]\{164\}$ &  
$89(129)[145]\{145\}$ &  
$100(143)[155]\{154\}$ \\
                             &  
$105(148)[164]\{164\}$  &  
$92(130)[145]\{145\}$  &  
$100(141)[154]\{155\}$  \\ \hline
\multicolumn{4}{|c|}
{\rule[0cm]{0cm}{0cm}
$M_{{\mathrm{had}}}>200$ GeV}
 \\ \hline  \hline
\multicolumn{4}{|c|}
{\rule[0cm]{0cm}{0cm}
$\sqrt s=400$ GeV}
 \\ \hline 
\multicolumn{4}{c}
{\rule{0cm}{1cm}
{\Large Tab. II}}  \\
\multicolumn{4}{c}
{\rule{0cm}{0cm}}
\end{tabular}
\end{center}
\end{table}

\clearpage

\begin{table}
\begin{center}
\begin{tabular}{|c|c|c|c|}
\hline
\rule[0cm]{0cm}{0cm}
$m_t$ (GeV)                           &\omit  
$~$                                   &\omit  
$\sigma(e^+e^-\ar X)$ (fb)            &
$~$                                   \\ \hline  \hline
\rule[0cm]{0cm}{0cm}
$~$                                   &  
$t\bar t$                             &  
$t\bar t \rightarrow \bar b bW^+W^-$  &  
$\bar b bW^+W^-$                      \\ \hline\hline
\rule[0cm]{0cm}{0cm}
$174$                        &  
$463(684)[698]\{698\}$  &  
$463(684)[698]\{698\}$  &  
$486(716)[725]\{725\}$  \\
                             &  
$429(682)[698]\{698\}$  &  
$429(681)[698]\{698\}$  &  
$451(713)[727]\{727\}$  \\ \hline
\rule[0cm]{0cm}{0cm}
$199$ &  
$372(537)[600]\{600\}$  &  
$365(518)[591]\{591\}$  &  
$386(549)[611]\{611\}$  \\
                             &  
$385(541)[600]\{600\}$  &  
$377(532)[591]\{591\}$  &  
$395(555)[613]\{611\}$  \\ \hline
\multicolumn{4}{|c|}
{\rule[0cm]{0cm}{0cm}
no cut}
 \\ \hline  \hline
\rule[0cm]{0cm}{0cm}
$174$ &  
$463(684)[698]\{698\}$ &  
$452(671)[685]\{685\}$ &  
$462(685)[696]\{697\}$ \\
                             &  
$429(681)[698]\{698\}$  &  
$419(669)[685]\{685\}$  &  
$428(683)[698]\{698\}$  \\ \hline
\rule[0cm]{0cm}{0cm}
$199$ &  
$372(537)[600]\{600\}$ &  
$354(516)[577]\{577\}$ &  
$365(527)[588]\{588\}$ \\
                             &  
$385(541)[600]\{600\}$  &  
$366(520)[577]\{577\}$  &  
$375(531)[588]\{588\}$  \\ \hline
\multicolumn{4}{|c|}
{\rule[0cm]{0cm}{0cm}
$0.95\le x_E\le1.05$}
 \\ \hline  \hline
\multicolumn{4}{|c|}
{\rule[0cm]{0cm}{0cm}
$\sqrt s=500$ GeV}
 \\ \hline 
\multicolumn{4}{c}
{\rule{0cm}{1cm}
{\Large Tab. III}}  \\
\multicolumn{4}{c}
{\rule{0cm}{0cm}}
\end{tabular}
\end{center}
\end{table}

\vfill
\newpage

\begin{figure}[p]~\epsfig{file=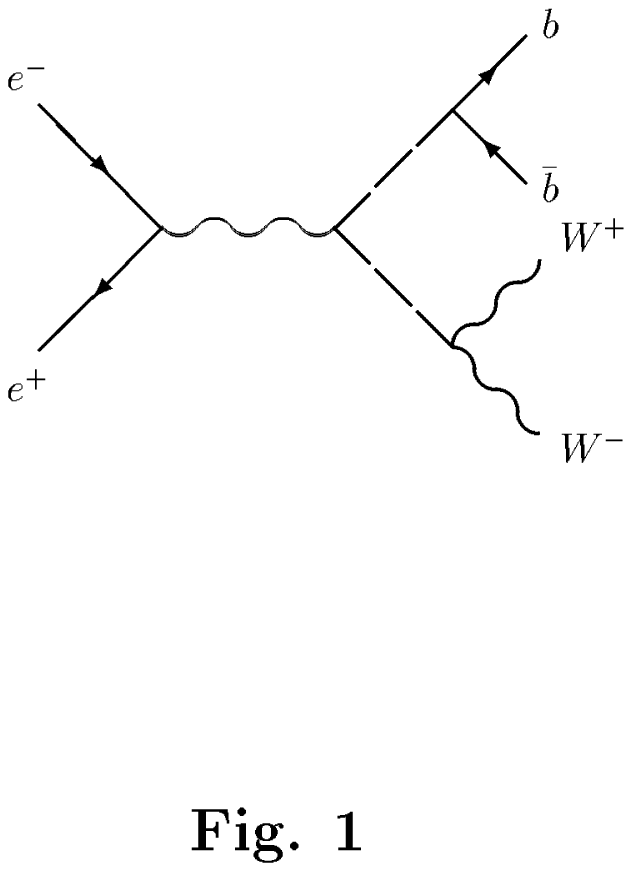,height=22cm}
\end{figure}
\stepcounter{figure}
\vfill
\clearpage

\begin{figure}[p]~\epsfig{file=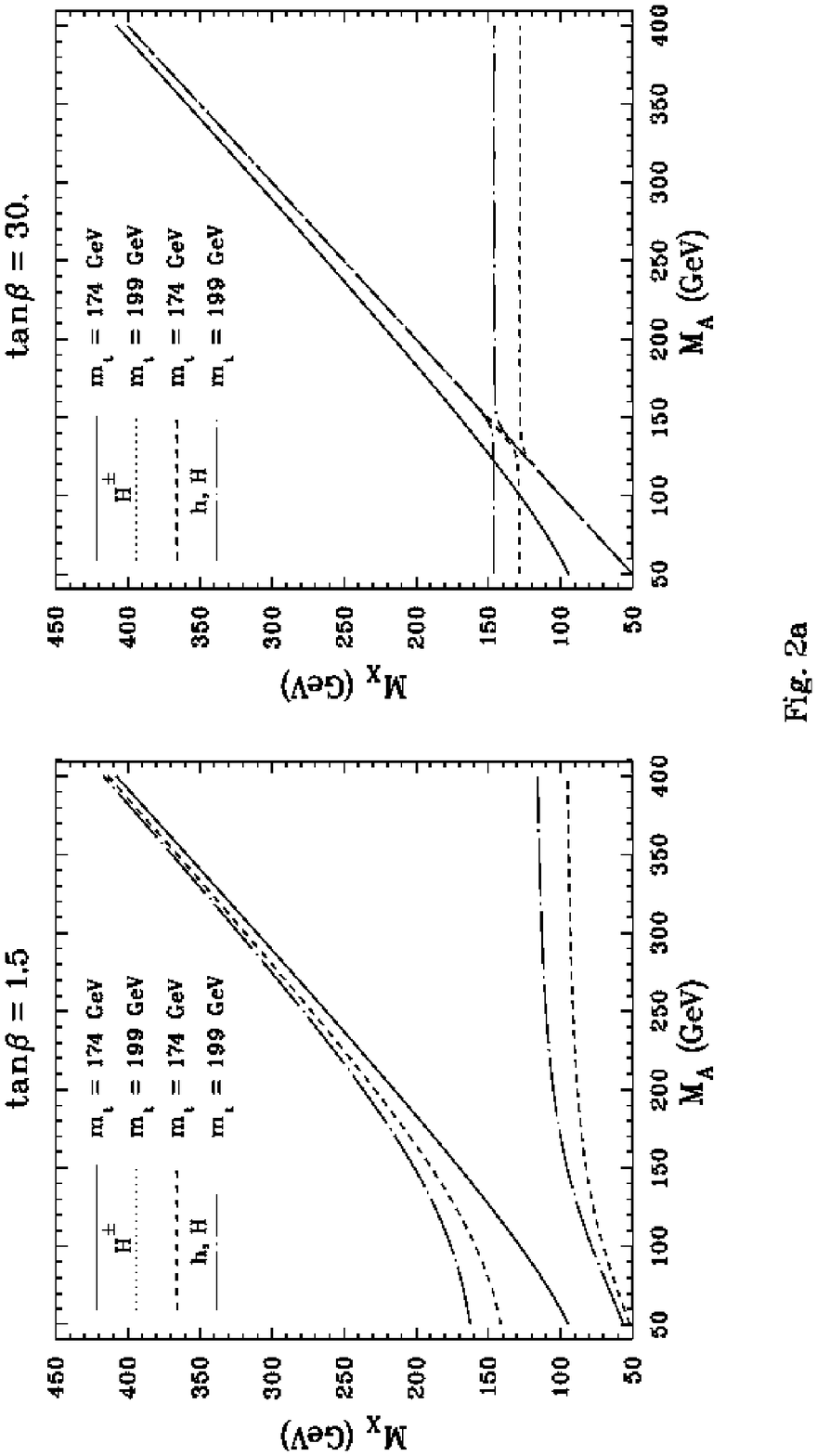,height=22cm}
\end{figure}
\stepcounter{figure}
\vfill
\clearpage

\begin{figure}[p]~\epsfig{file=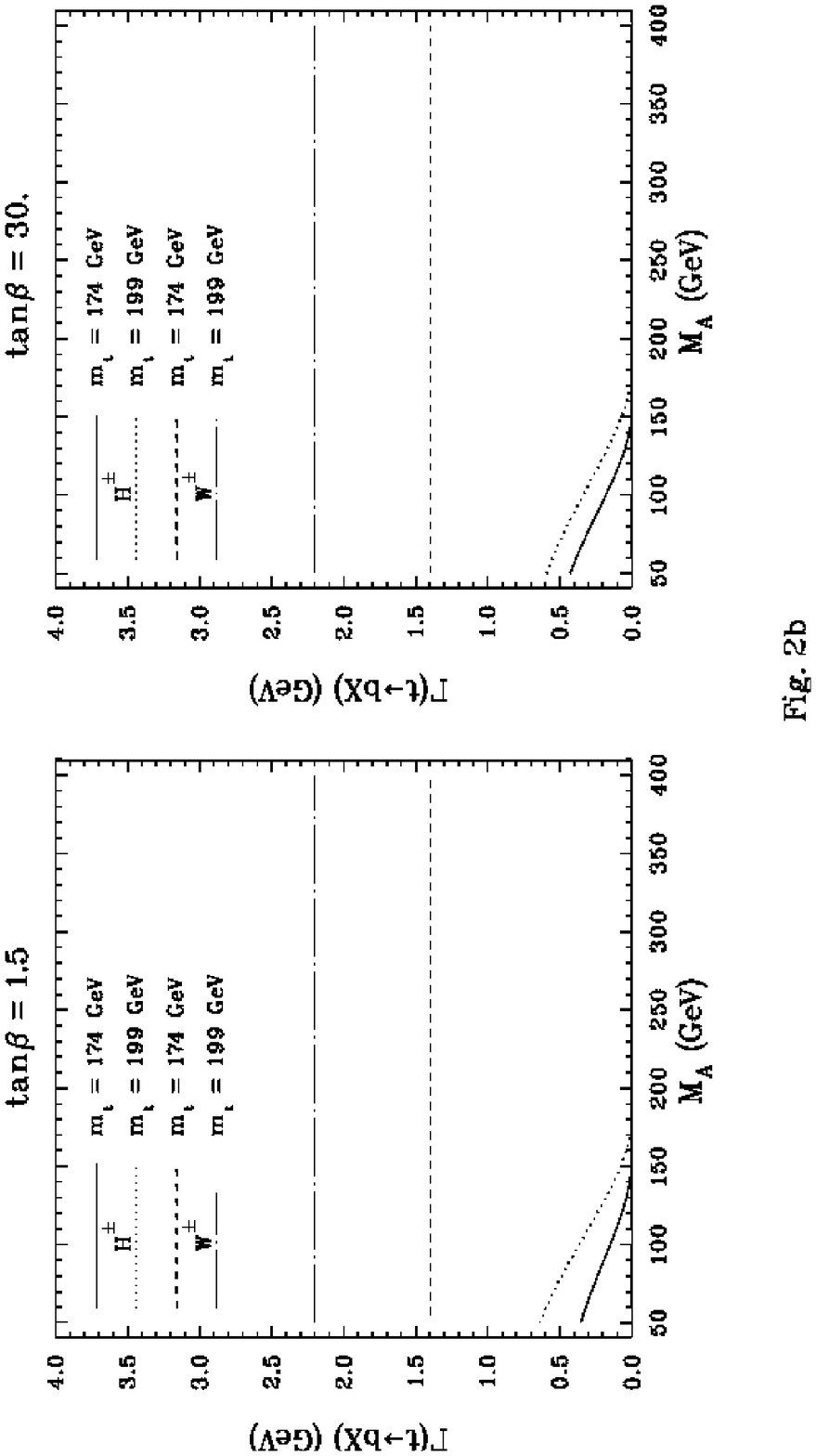,height=22cm}
\end{figure}
\stepcounter{figure}
\vfill
\clearpage

\begin{figure}[p]~\epsfig{file=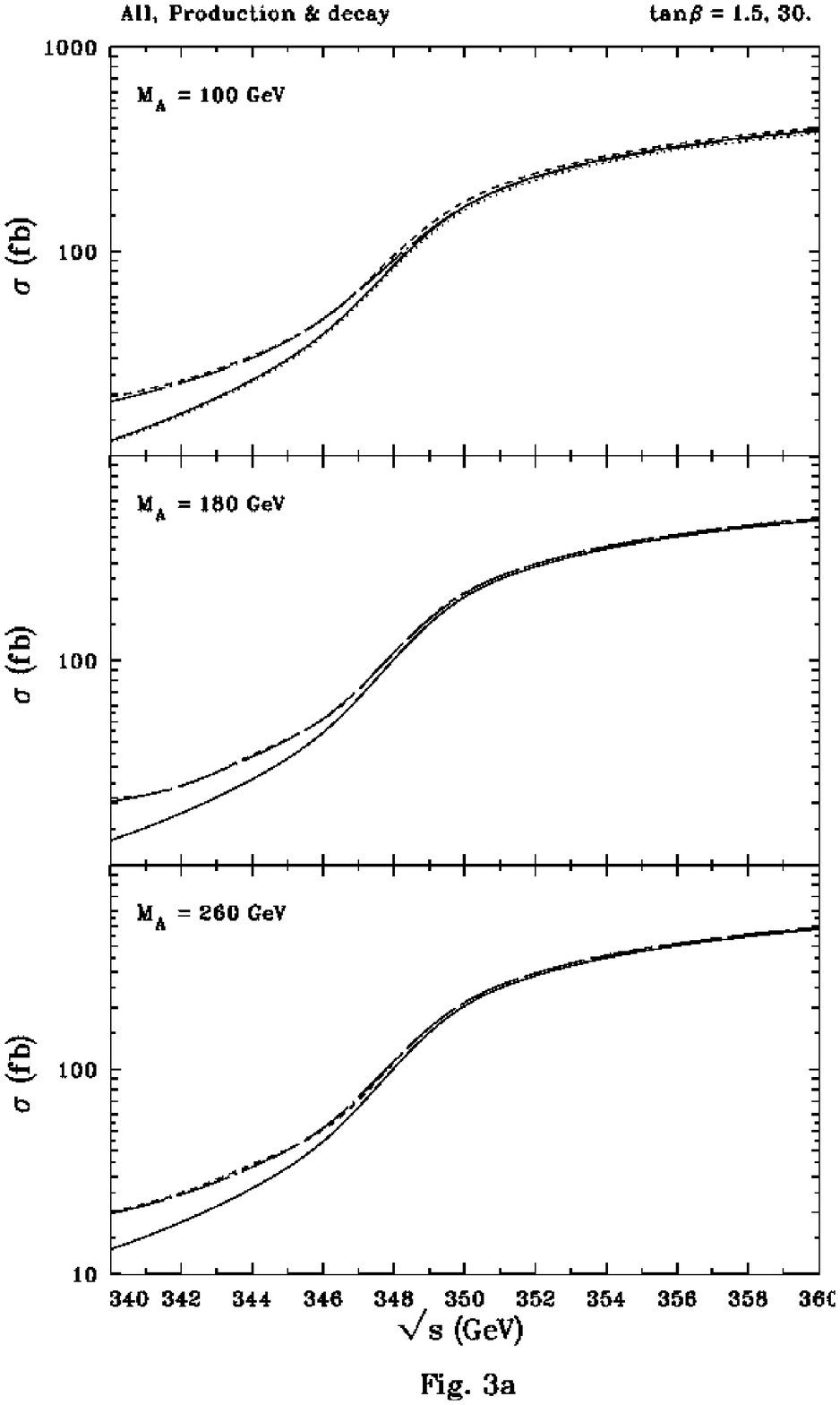,height=22cm}
\end{figure}
\stepcounter{figure}
\vfill
\clearpage

\begin{figure}[p]~\epsfig{file=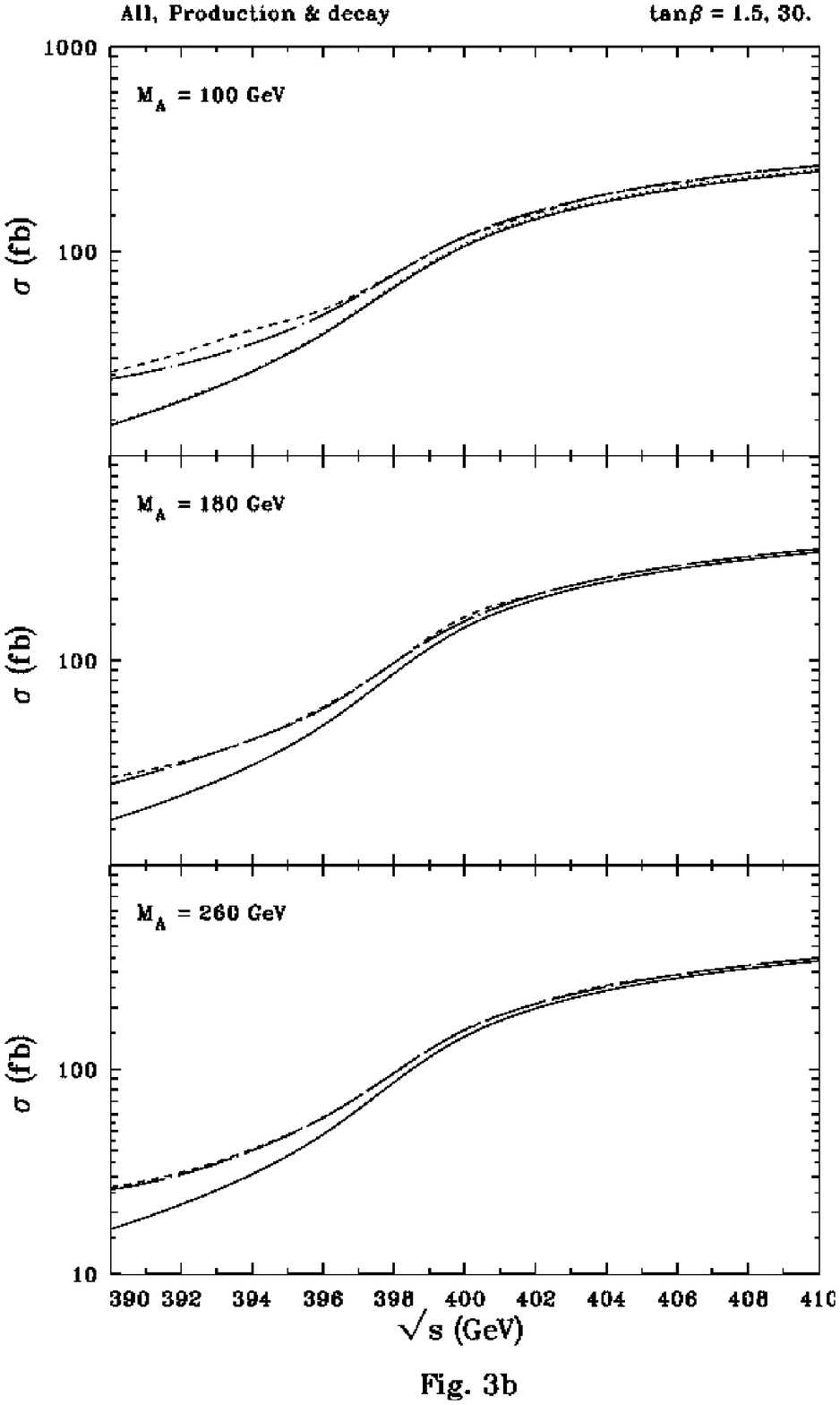,height=22cm}
\end{figure}
\stepcounter{figure}
\vfill
\clearpage

\begin{figure}[p]~\epsfig{file=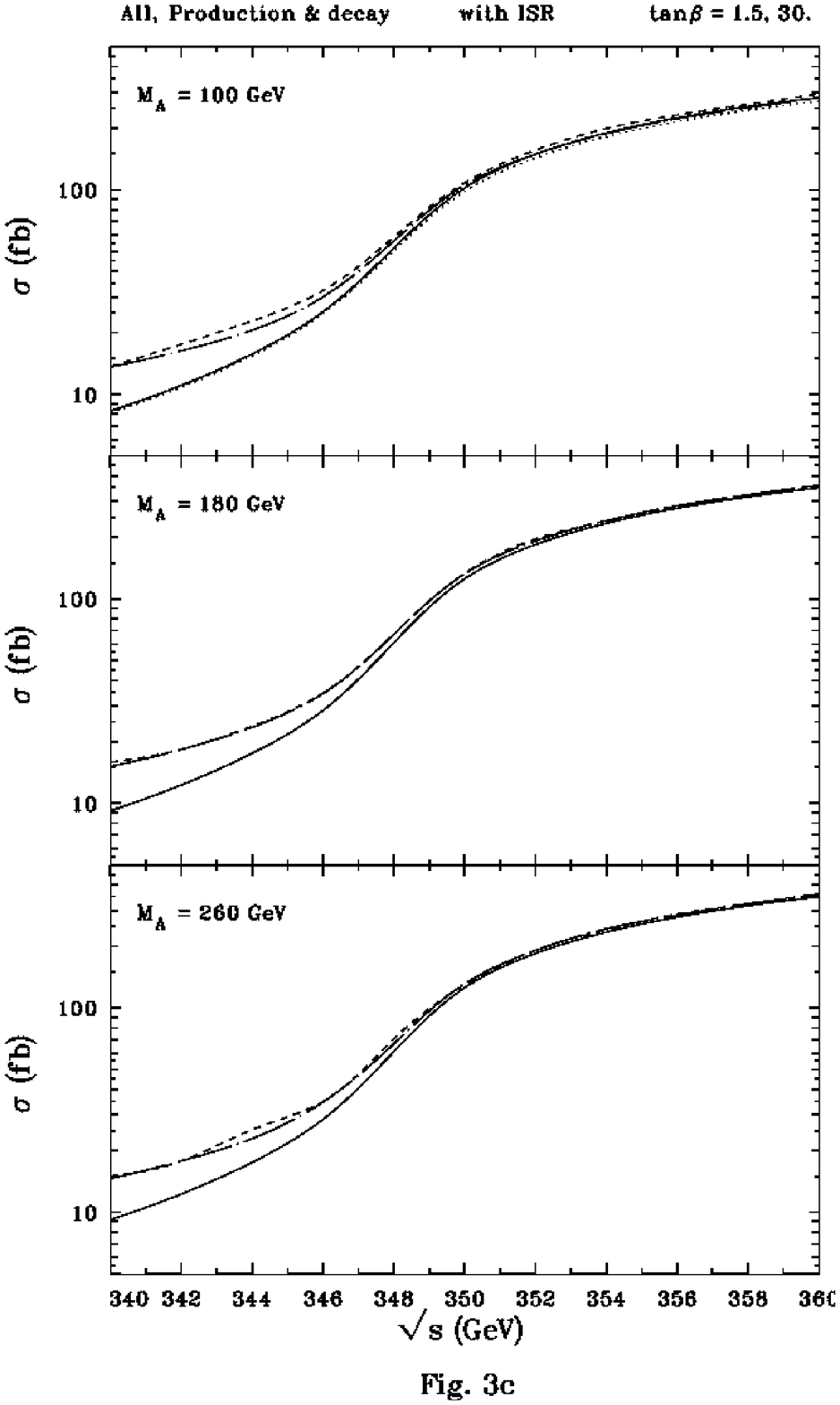,height=22cm}
\end{figure}
\stepcounter{figure}
\vfill
\clearpage

\begin{figure}[p]~\epsfig{file=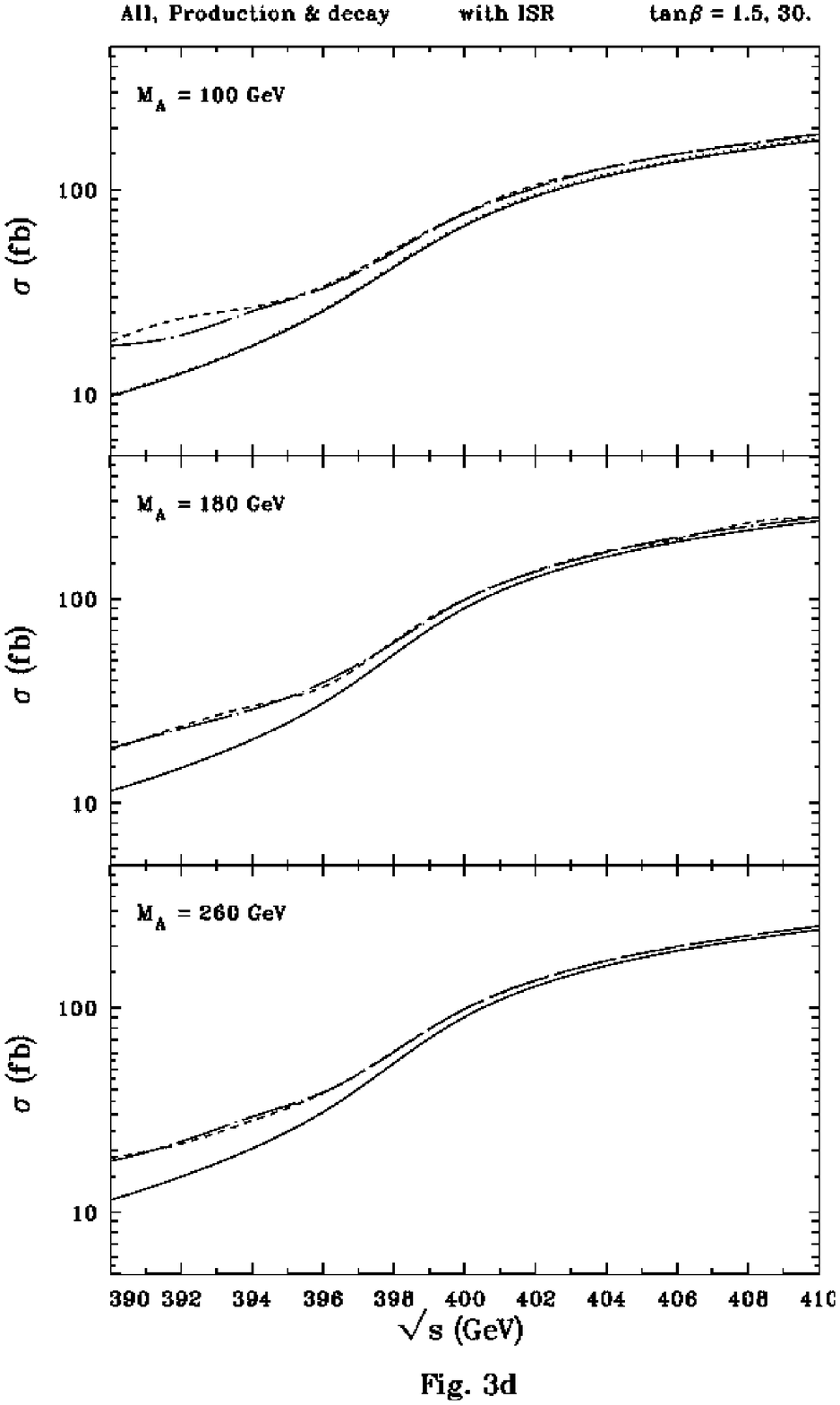,height=22cm}
\end{figure}
\stepcounter{figure}
\vfill
\clearpage

\begin{figure}[p]~\epsfig{file=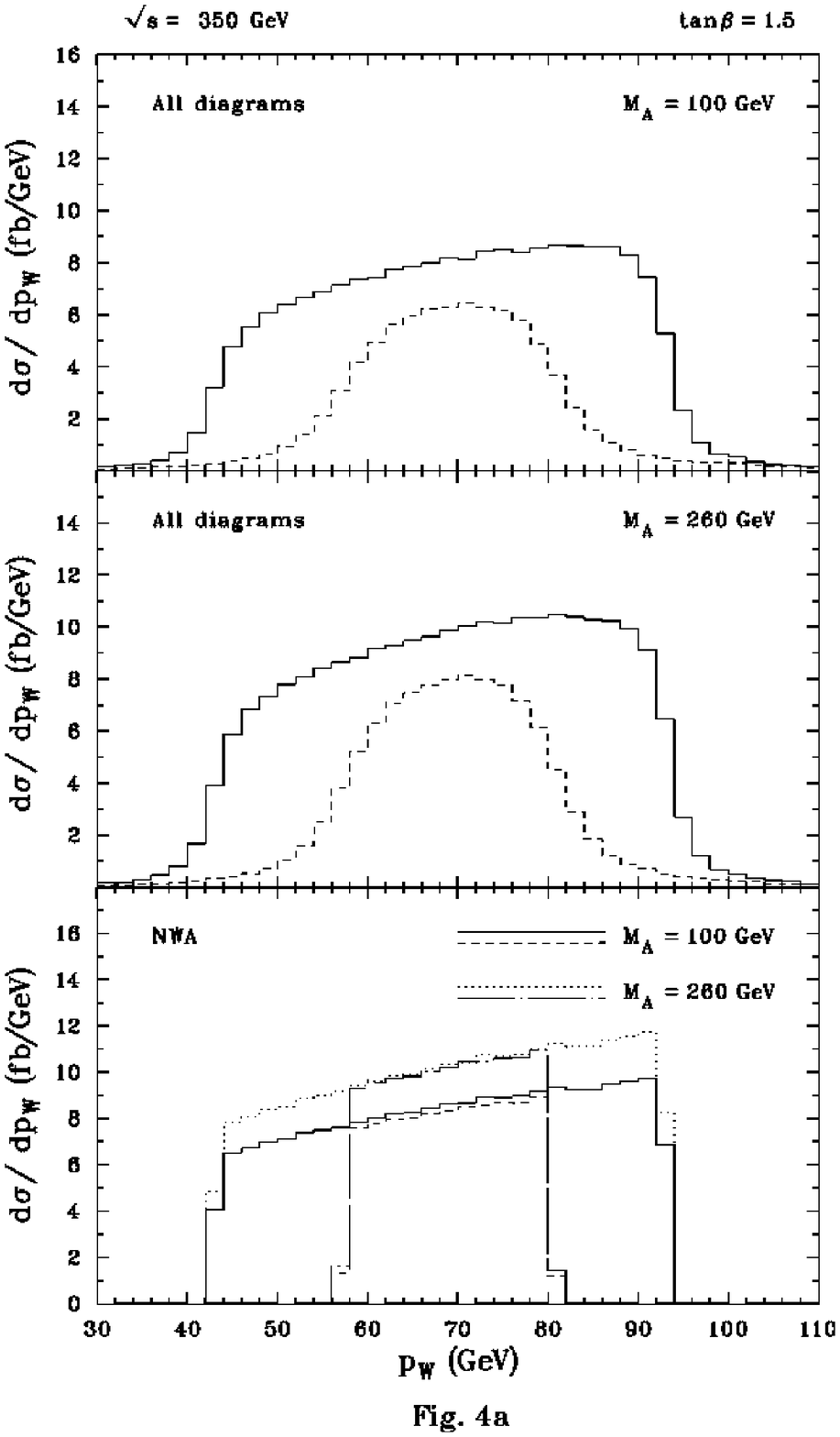,height=22cm}
\end{figure}
\stepcounter{figure}
\vfill
\clearpage

\begin{figure}[p]~\epsfig{file=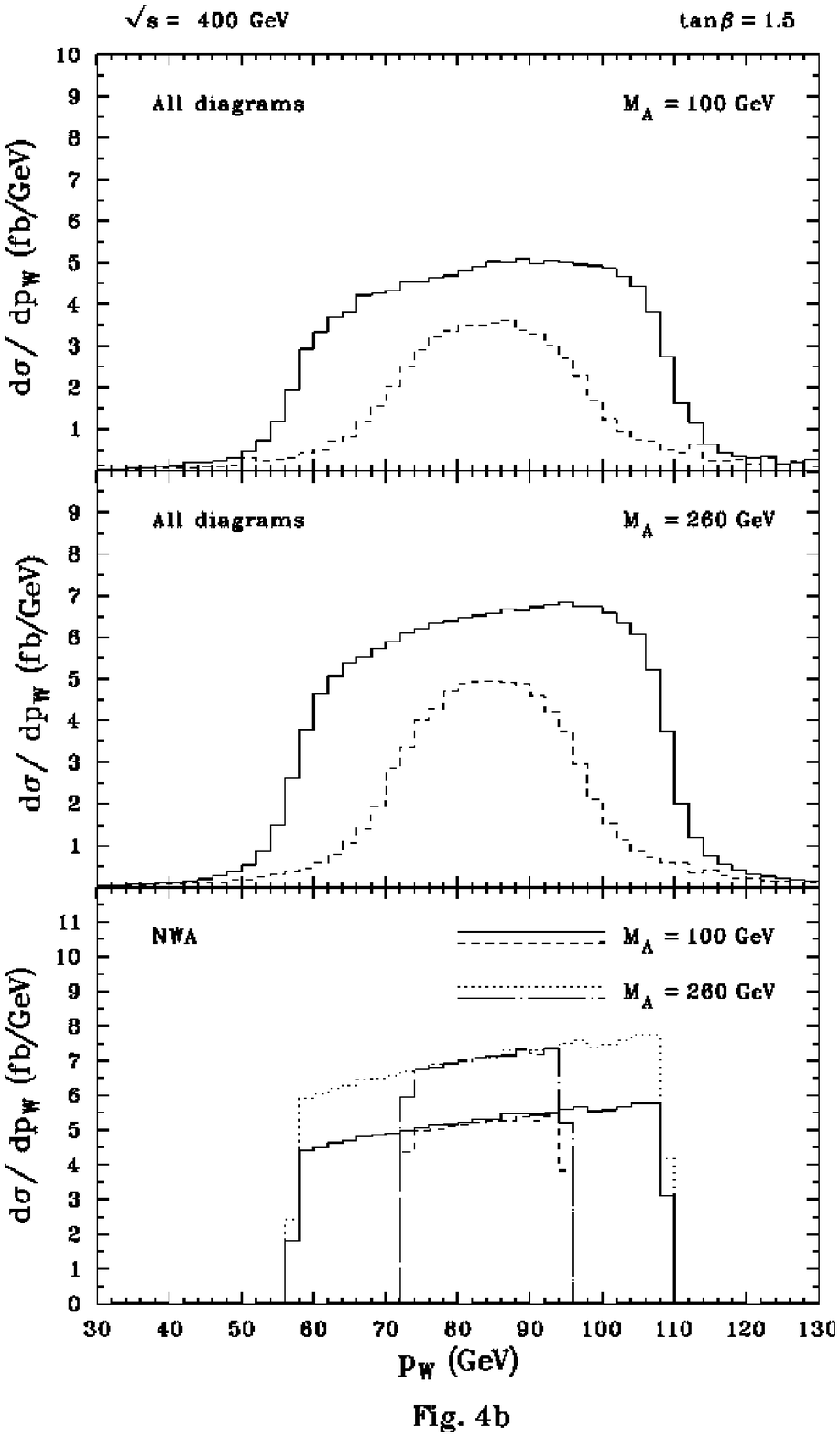,height=22cm}
\end{figure}
\stepcounter{figure}
\vfill
\clearpage

\begin{figure}[p]~\epsfig{file=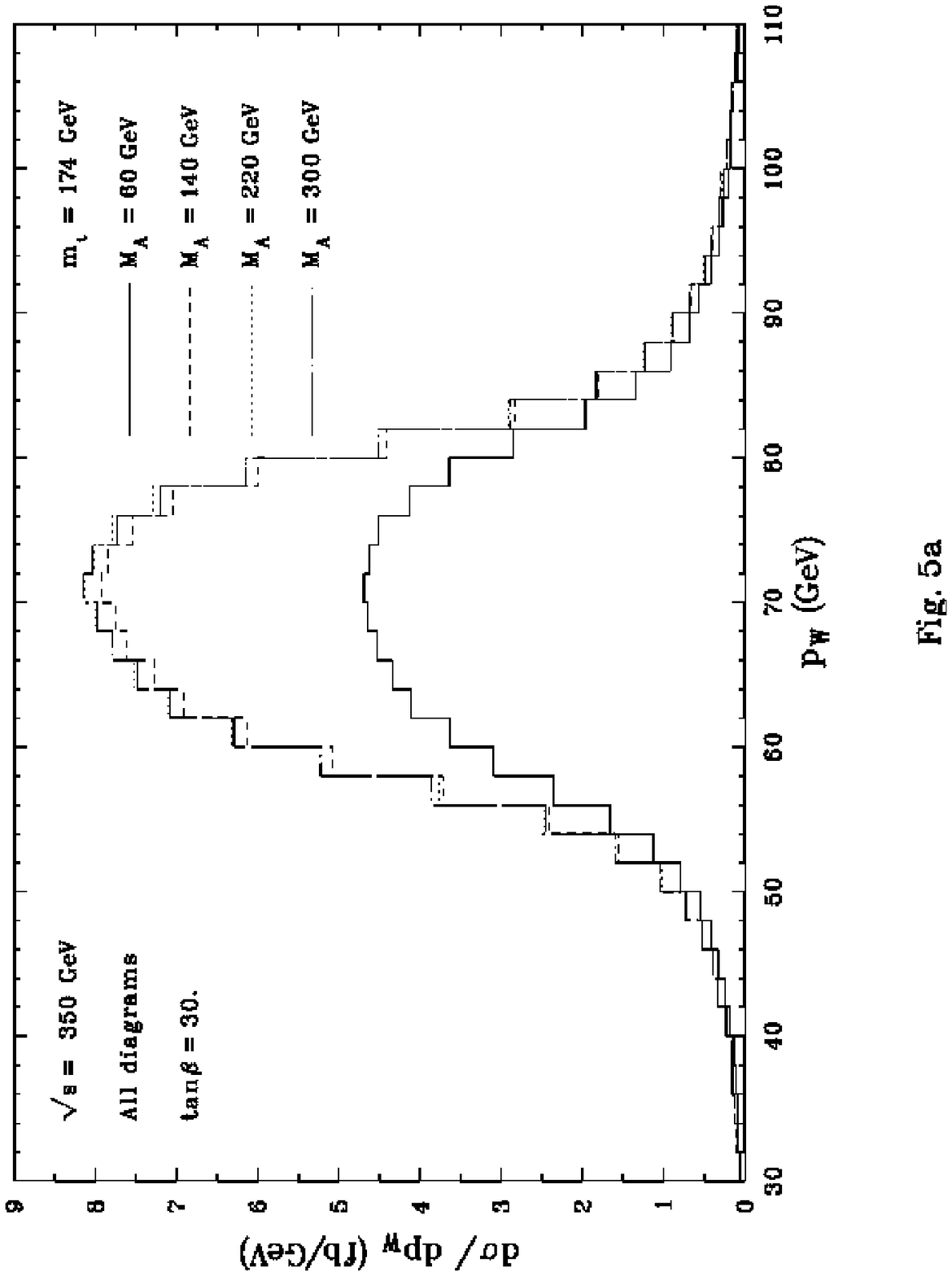,height=22cm}
\end{figure}
\stepcounter{figure}
\vfill
\clearpage

\begin{figure}[p]~\epsfig{file=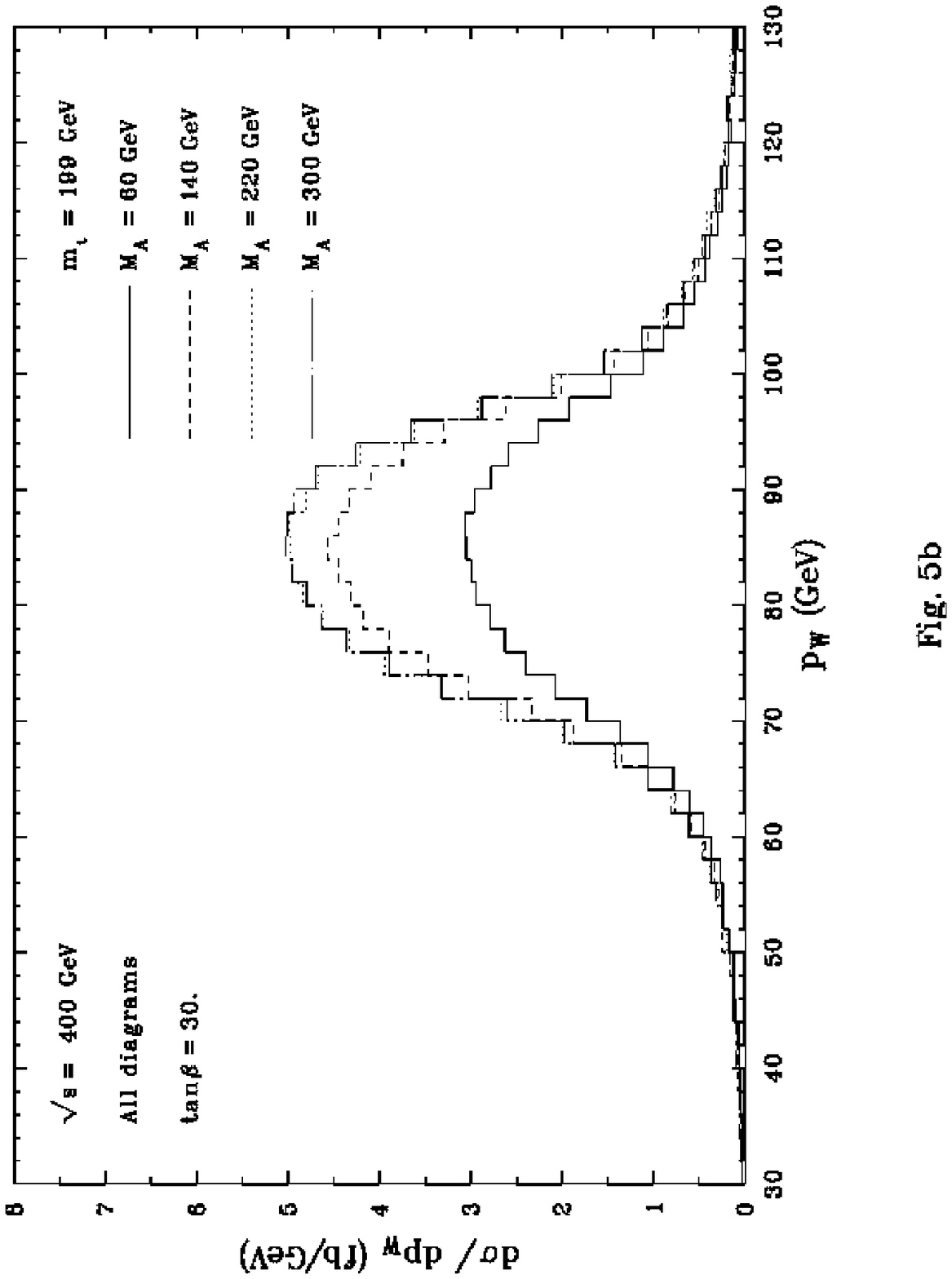,height=22cm}
\end{figure}
\stepcounter{figure}
\vfill
\clearpage

\begin{figure}[p]~\epsfig{file=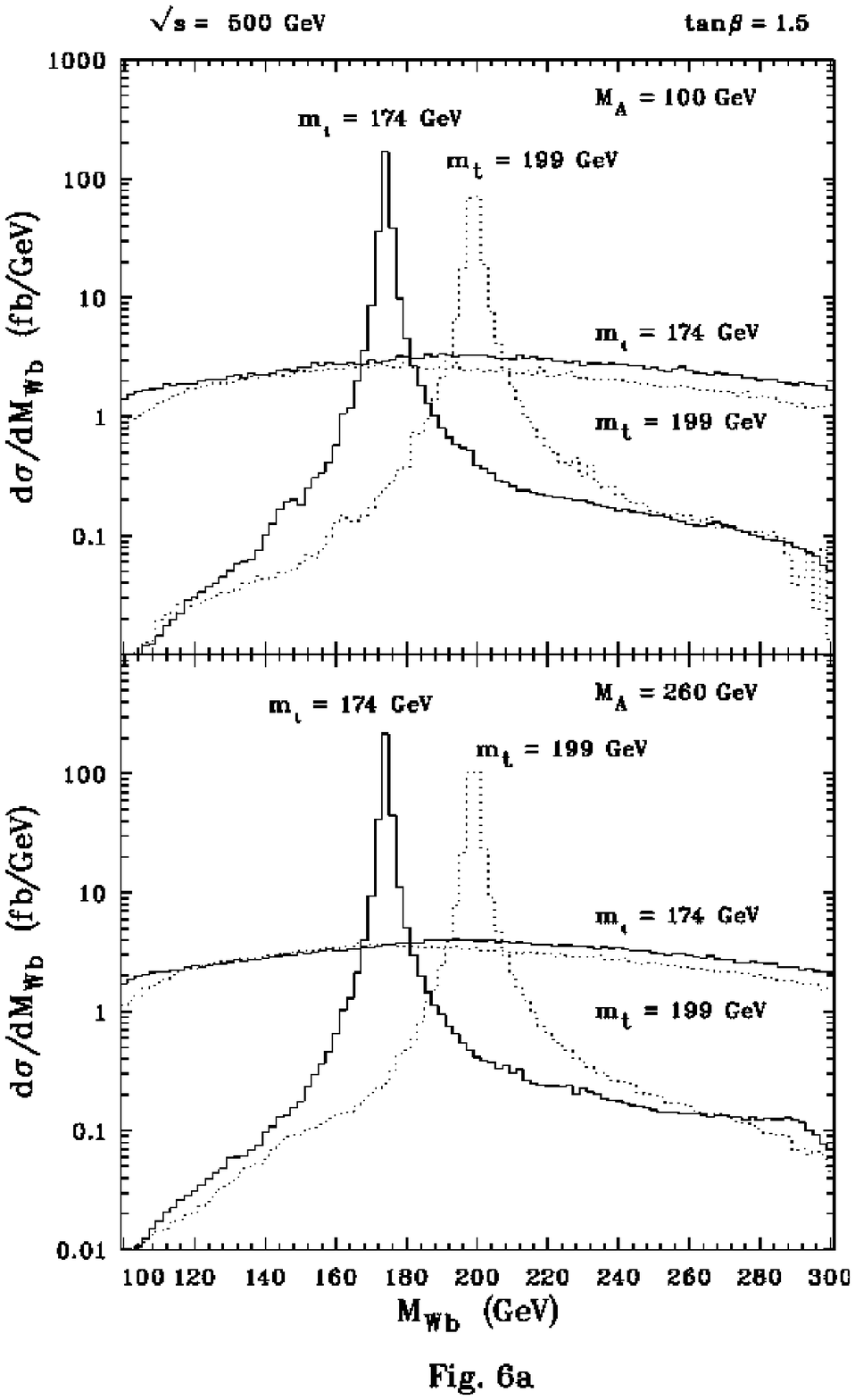,height=22cm}
\end{figure}
\stepcounter{figure}
\vfill
\clearpage

\begin{figure}[p]~\epsfig{file=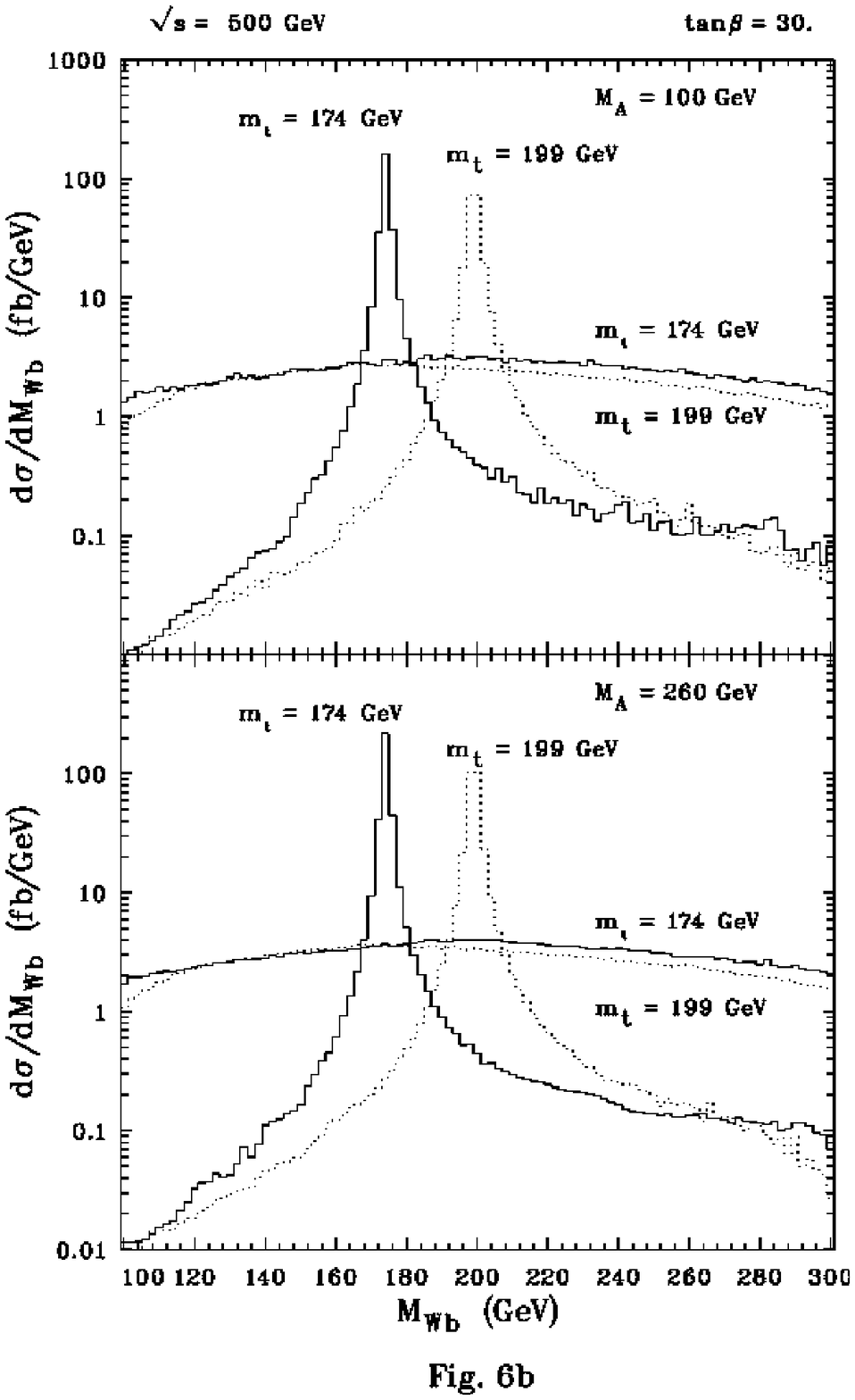,height=22cm}
\end{figure}
\stepcounter{figure}
\vfill
\clearpage

\begin{figure}[p]~\epsfig{file=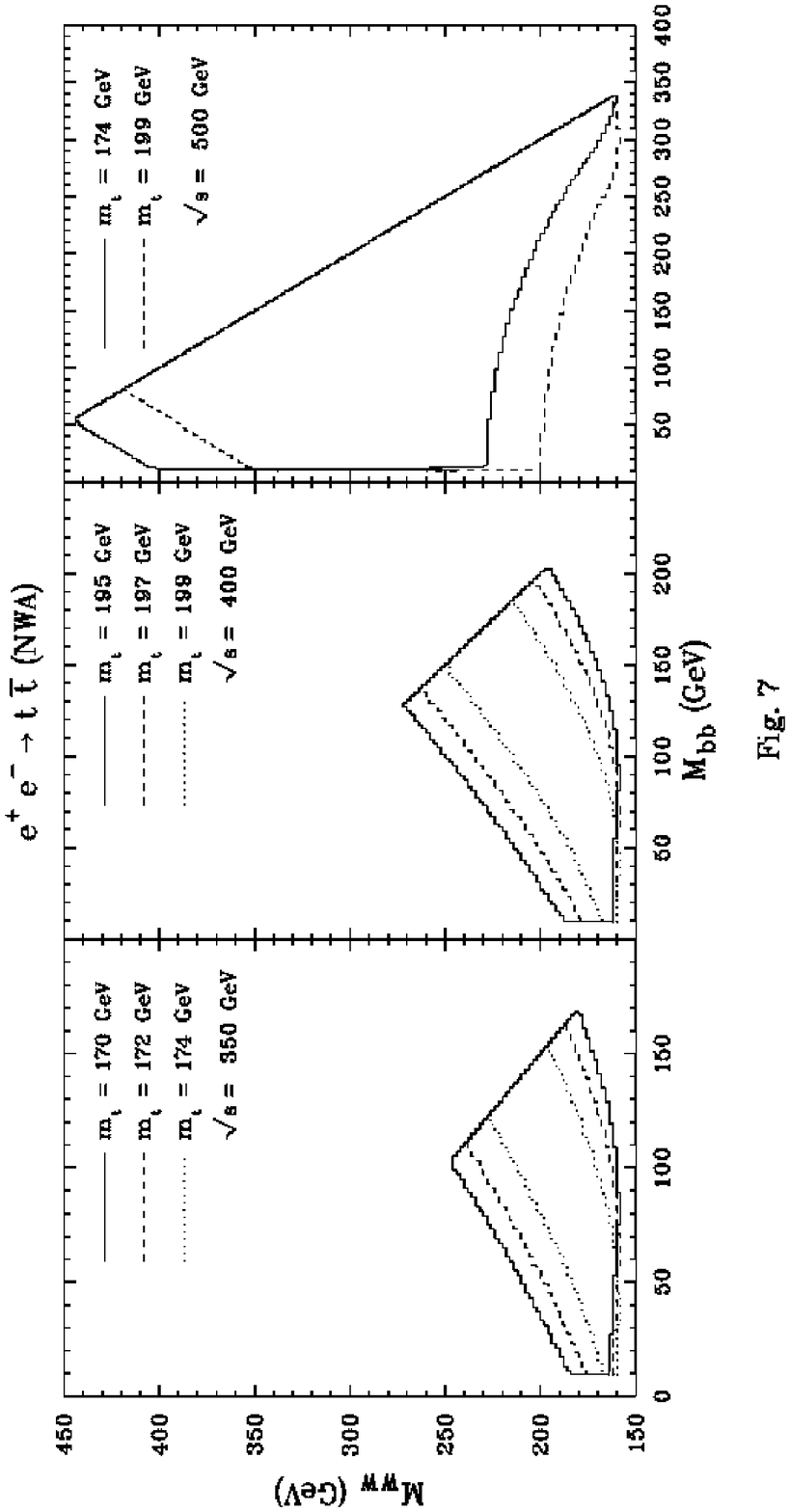,height=22cm}
\end{figure}
\stepcounter{figure}
\vfill
\clearpage

\begin{figure}[p]~\epsfig{file=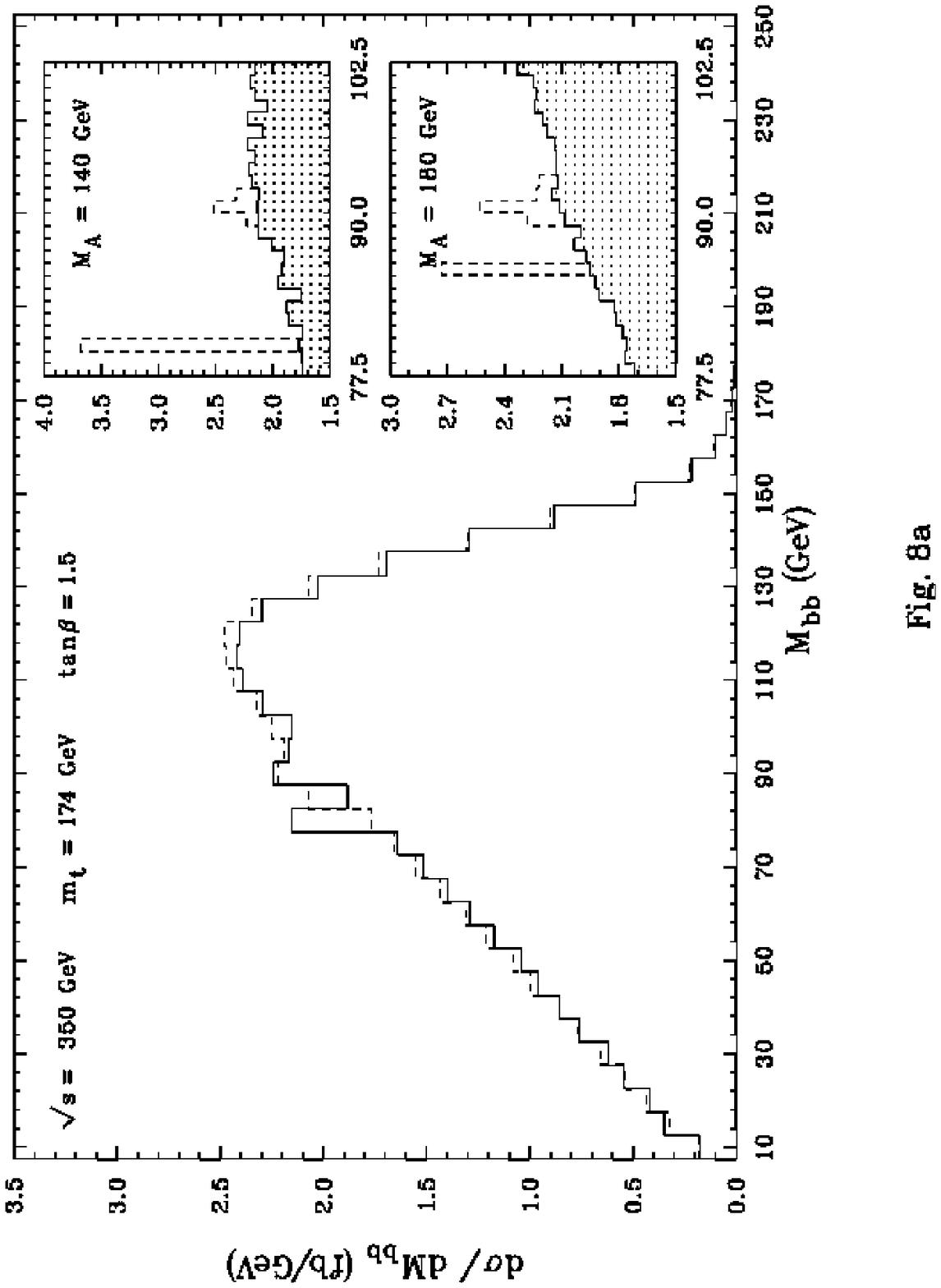,height=22cm}
\end{figure}
\stepcounter{figure}
\vfill
\clearpage

\begin{figure}[p]~\epsfig{file=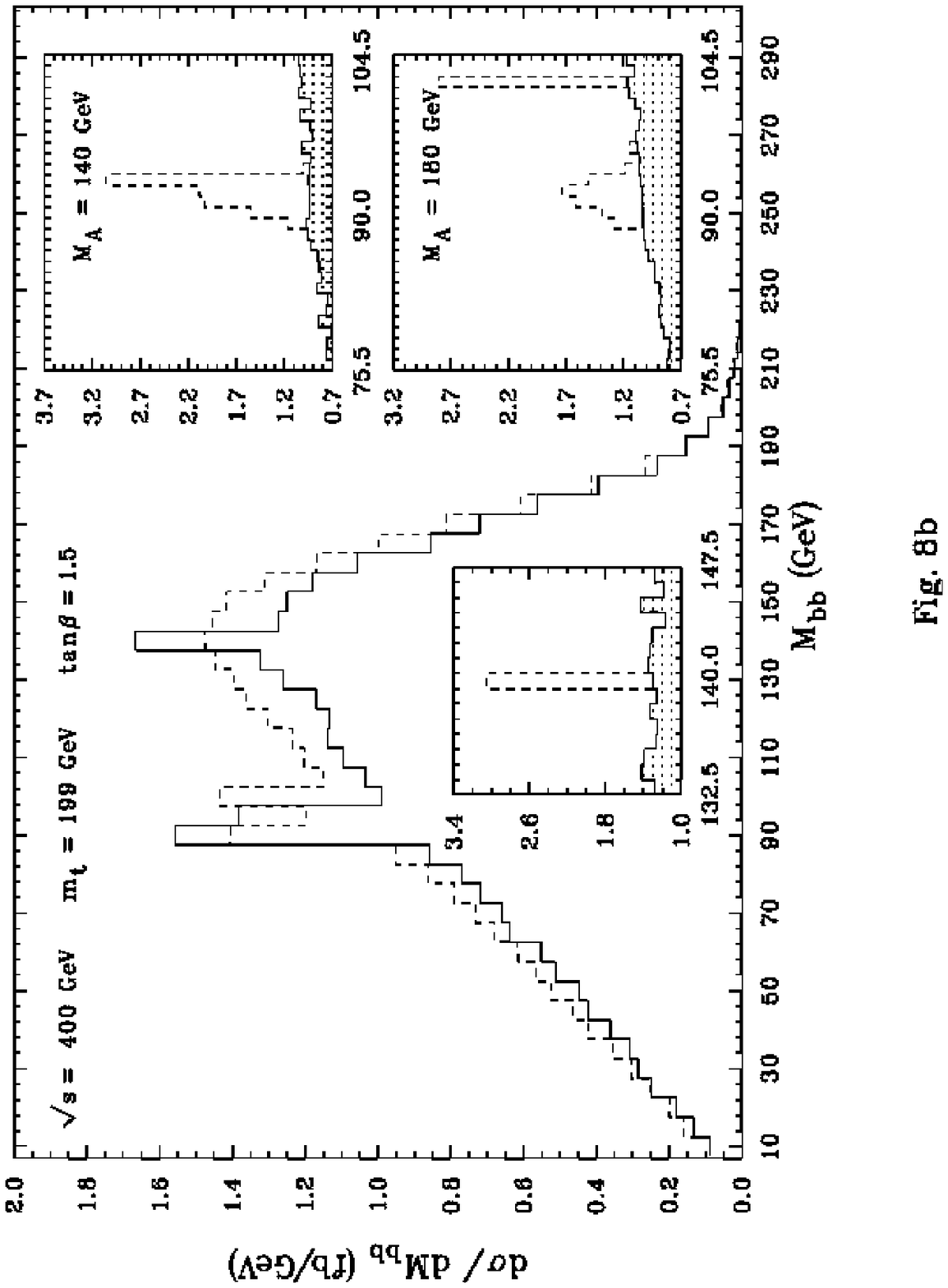,height=22cm}
\end{figure}
\stepcounter{figure}
\vfill
\clearpage

\begin{figure}[p]~\epsfig{file=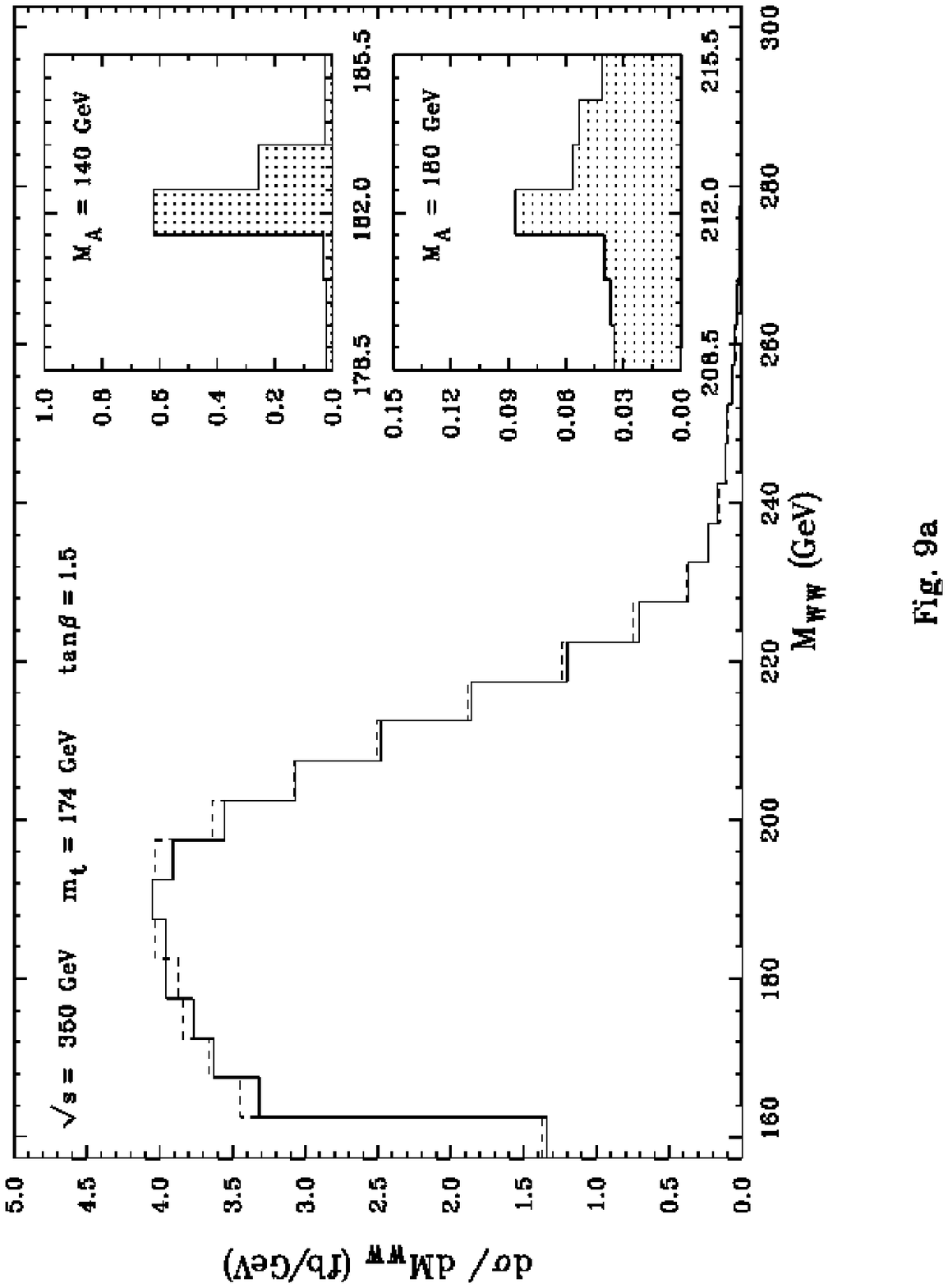,height=22cm}
\end{figure}
\stepcounter{figure}
\vfill
\clearpage

\begin{figure}[p]~\epsfig{file=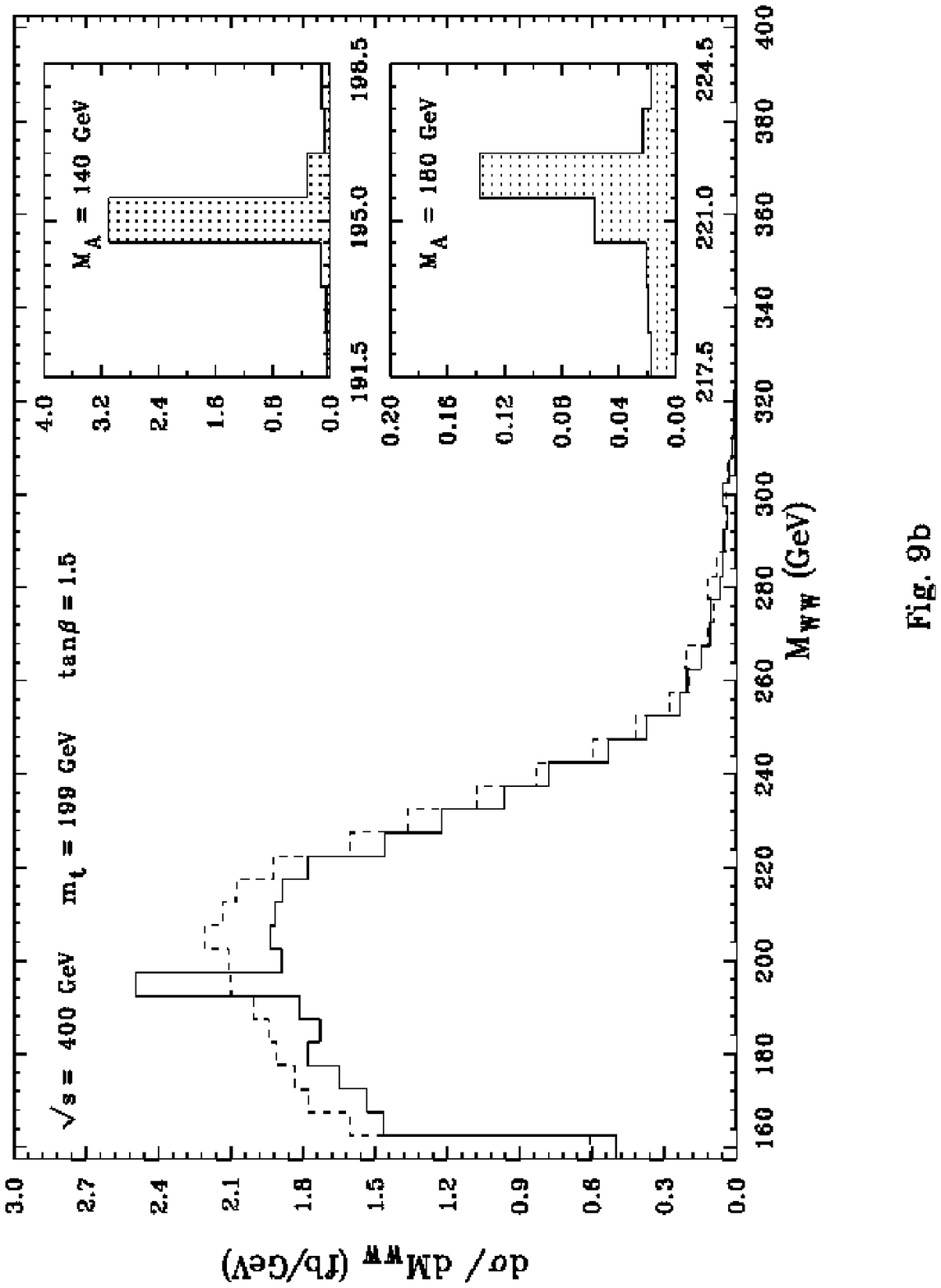,height=22cm}
\end{figure}
\stepcounter{figure}
\vfill
\clearpage

\begin{figure}[p]~\epsfig{file=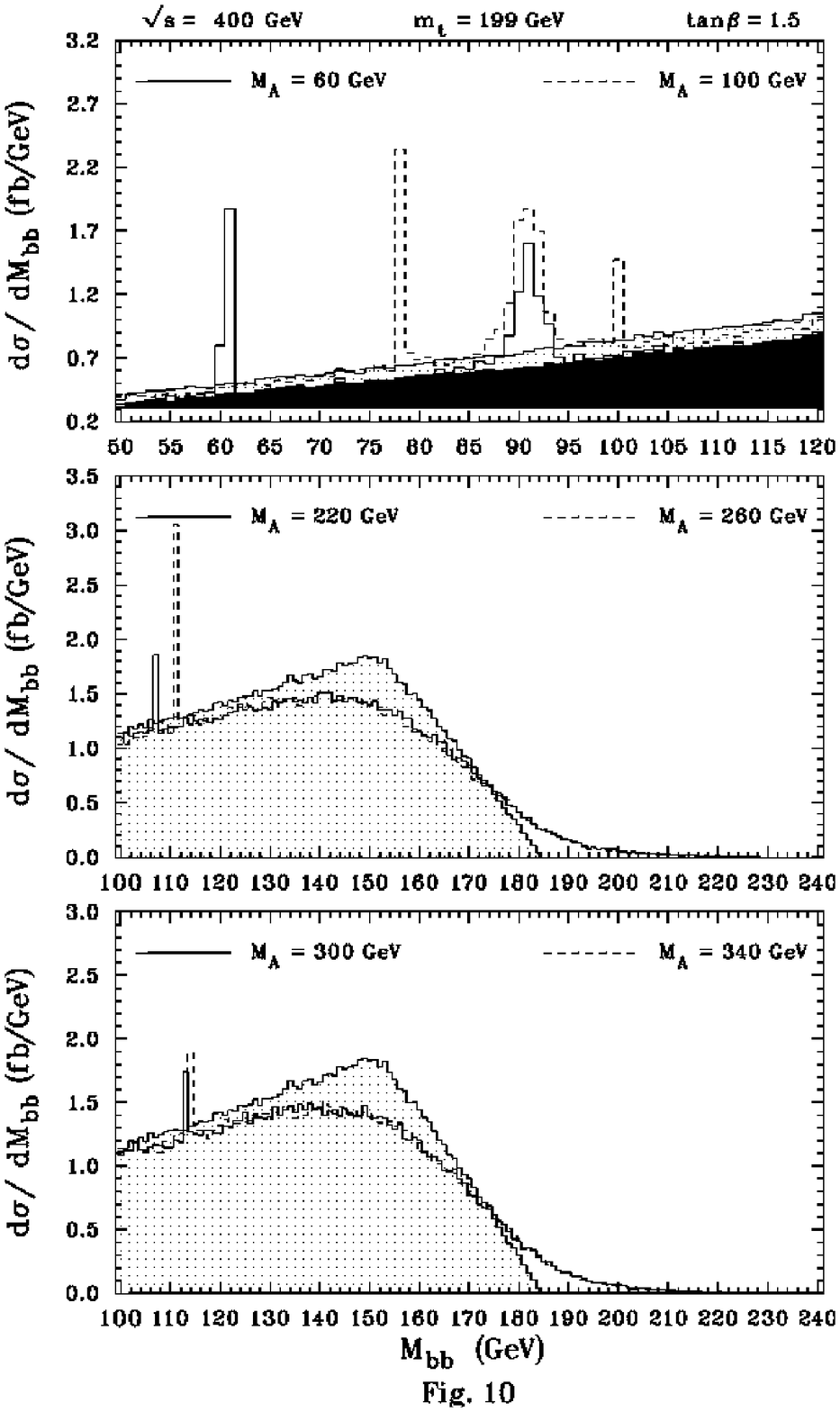,height=22cm}
\end{figure}
\stepcounter{figure}
\vfill
\clearpage

\begin{figure}[p]~\epsfig{file=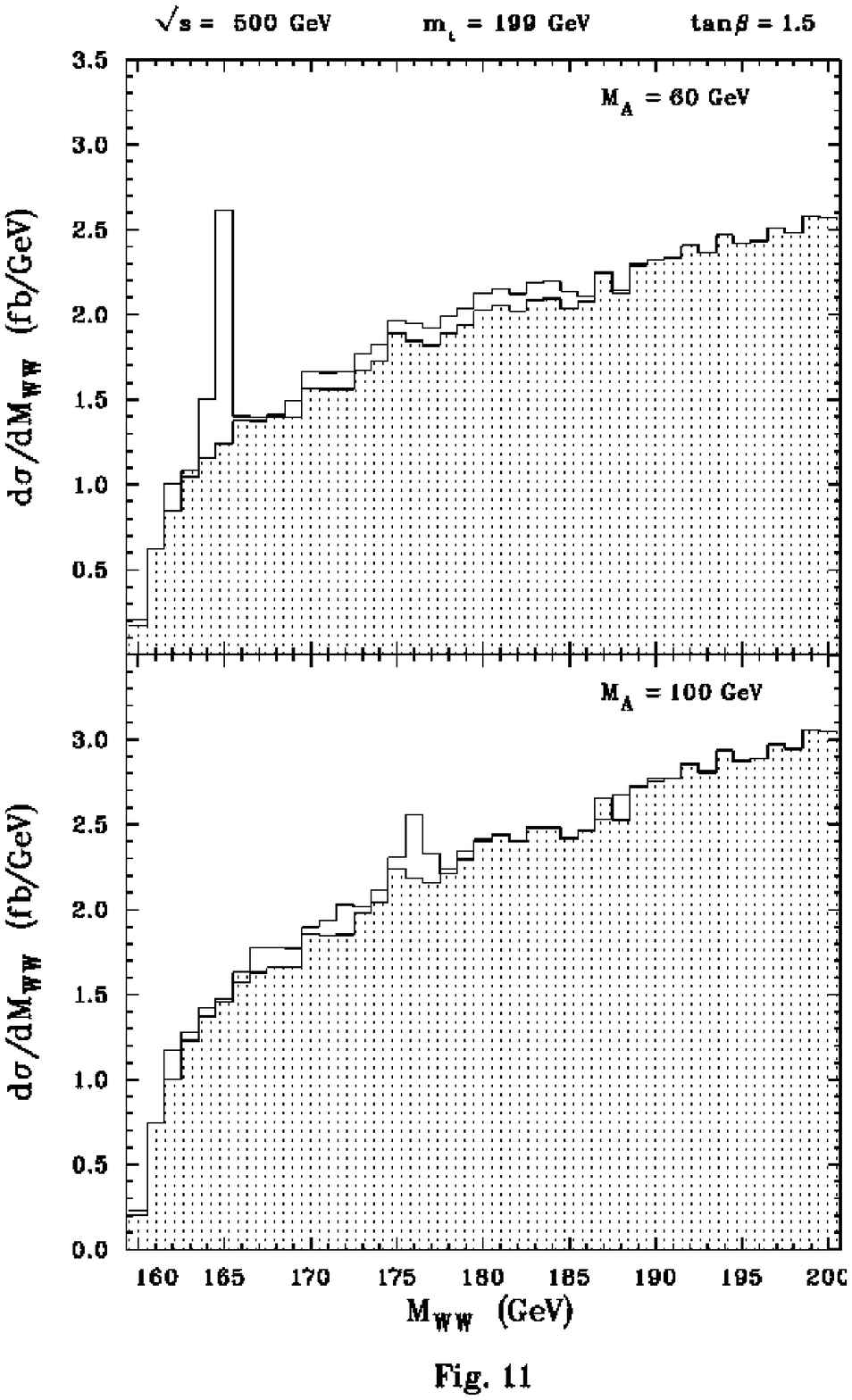,height=22cm}
\end{figure}
\stepcounter{figure}
\vfill
\clearpage


\begin{thebibliography}{99}

\bibitem{ee500} Proceedings of the Workshop ``{\it $e^+e^-$ Collisions at 
500 GeV. The Physics Potential}\ '', 
Munich, Annecy, Hamburg, 3--4 February 1991, ed. P.M.~Zerwas, DESY 92--123A/B,
August 1992, DESY 93--123C, December 1993.

\bibitem{Teva} Proceedings of the Workshop on ``{\it Physics at Current
Accelerators and Supercolliders}\ '', eds. J. Hewett, A. White and 
D. Zeppenfeld,
Argonne National Laboratory, 1993.

\bibitem{lep2w} G. Altarelli, T. Sj\"ostrand and F. Zwirner, eds.,
`{\it Report of the Workshop  on  Physics at LEP2}', CERN 96--01 (1996).

\bibitem{CMS}  CMS Technical Proposal, CERN/LHC/94-43 LHCC/P1, December 1994.

\bibitem{ATLAS} ATLAS Technical Proposal,
CERN/LHC/94-43 LHCC/P2, December 1994.

\bibitem{tt} A. Ballestrero, E. Maina and S. Moretti, \pl B333 1994 434.

\bibitem{ZH} A. Ballestrero, E. Maina and S. Moretti, \pl B335 1994 460.

\bibitem{ideal} Proceedings of the Workshop on ``{\it High Luminosities
at LEP}\ '', CERN Report 91-02, Geneva, Switzerland.

\bibitem{Guasch} J. Guasch and J. Sol\`a, \preprint\ UAB-FT-389, March 1996.

\bibitem{bern} W. Bernreuther {\it et al.}, in Ref.~\cite{ee500}, part A.

\bibitem{ref30b} C.S. Li and T.C. Yuan, \pr D42 1990 3088. 

\bibitem{ref27} A. Denner, R.J. Guth and J.H.~K\"uhn, 
{\it Nucl. Phys.} {\bf B377} (1992) 3.

\bibitem{proceed} A. Djouadi, J. Kalinowski and P.M. Zerwas, in 
Ref.~\cite{ee500}, part A.

\bibitem{DKZ} A. Djouadi, J. Kalinowski and P.M. Zerwas, in
Ref.~\cite{ee500}, part A (and references therein).

\bibitem{gattoevolpe} A. Ballestrero and E. Maina, {\it
Phys. Lett.} {\bf B350} (1995) 225.

\bibitem{HZ} K.~Hagiwara and D.~Zeppenfeld,
{\it Nucl. Phys.} {\bf B274} (1986) 1. 

\bibitem{ioejames} S. Moretti and W.J. Stirling, {\it Phys. Lett.} 
{\bf B347} (1995) 291; Erratum, {\it ibidem} {\bf B366} (1996) 451.

\bibitem{Alimit} See for example:\\
D. Treille, in Proceedings of the Workshop on ``{\it Physics
and Experiments with Linear $e^+e^-$ Colliders}\ '', Waikoloa, Hawaii,
1993, eds. F. Harris {\it et al.}, Vol.~I (and references therein).

\bibitem{widthtopSM1} R. Kleiss and W.J. Stirling, \zp C40 1988 419.

\bibitem{widthtopSM2} I.~Bigi, Y. Dokshitzer, V.A. Khoze, J. K\"uhn 
and P.M. Zerwas, \pl B181 1986 157.

\bibitem{widthtopMSSM} G.L.~Kane, 
Proceedings of the ``{\it Madison Workshop}\ '' (1979).

\bibitem{CDFtop} CDF~Collaboration,
{\it Phys. Rev. Lett.} {\bf 74} (1995) 2626.

\bibitem{D0top} D0 Collaboration, {\it Phys. Rev. Lett.} {\bf 74} (1995) 2632.

\bibitem{newtop} A. Caner, Presented at ``{\it Rencontres du Physique de la 
Valle d'Aoste}'', March 1996;\\
M. Narain, Presented at the `{\it Rencontres du Physique de 
la Valle d'Aoste}'',
March 1996. 

\bibitem{reviewtop} M. Je\.zabek and J.H. K\"uhn, \preprint\ TTP-93-4,
February 1993.

\bibitem{orange54} J. Jersak, E. Laermann and P.M. Zerwas, \pr D25 1982 363;\\
J. Schwinger, ``Particles, Sources and Fields''
(Addison-Wesley, Reading MA, 1973);\\
L. Reinders, H. Rubinstein and S. Yazaki, \prep C127 1985 1;\\
S. G\"usken, J.H. K\"uhn and P.M. Zerwas, \pl B155 1985 185. 

\bibitem{vak} V.S. Fadin and V.A. Khoze, {\it JETP Lett.} {\bf 46} (1987) 
525;
{\it Sov. J. Nucl. Phys.} {\bf 53} (1988) 692.

\bibitem{ref9} M.J. Strassler and M.E. Peskin, \pr D43 1991 1500.

\bibitem{thr} V.S.~Fadin and O.I.~Yakovlev, 
{\it Sov. J. Nucl. Phys.} {\bf 23} (1991) 1117;\\
M.~Je\.zabek, J.H.~K\"uhn and T.~Teubner,
{\it Z. Phys.} {\bf C56} (1992) 653;\\
Y.~Sumino, K.~Fujii, K.~Hagiwara, H.~Murayama and C.-K.~Ng,
\pr D47 1992 56;\\
M.~Je\.zabek and T.~Teubner, \zp C59 1993 669.

\bibitem{orange58} A. Denner and T. Sack, \np B348 1991 46;\\
W. Beenakker, S.C. van der Marck and W. Hollik,
\np B365 1991 24.

\bibitem{newcorr} V. Driesen, W. Hollik and A. Kraft, \preprint\ KA-TP-1996,
March 1996, to appear in 
Proceedings of the Workshop ``{\it Physics with $e^+e^-$ Colliders}\ '', 
Annecy, Gran Sasso, Hamburg, 1995.

\bibitem{Bagliesi} G. Bagliesi {\it et al.}, in Ref.~\cite{ee500}, part A.

\bibitem{Poland} 
P. Chankowski, S. Pokorski and J. Rosiek, {\it Phys. Lett.} 
{\bf B274} (1992) 191; {\it Phys. Lett.} {\bf B286} (1993) 307;
{\it Nucl. Phys.} {\bf B423} (1994) 437 and 497;\\
J. Rosiek, A. Sopczak, P. Chankowski and S. Pokorski,
in  Ref.~\cite{ee500}, part C.

\bibitem{Dabel} A. Dabelstein and W. Hollik, in Ref.~\cite{ee500}, part C.

\bibitem{genuine} M. B\"ohm, A. Denner, T. Sack, W. Beenakker, F.A.
Berends and H. Kuijf, \np B304 1988 463;\\
J. Fleisher F. Jegerlehner and M. Zralek, \zp C42 1989 409.

\bibitem{expre} J.F.~Gunion, L. Roszkowski, A. Turski, H. Haber, G. Gamberini,
B. Kayser, S. Novaes, F. Olness and J. Wudka, \pr D38 1988 3444.

\bibitem{Janot} P. Janot, in Ref.~\cite{ee500}, part A
 (and references therein).

\bibitem{WWh} V. Barger, T. Han and A. Stange, \pr D42 1990 777;\\
              M.~Baillargeon, F.~Boudjema, F.~Cuypers, E.~Gabrielli and
B.~Mele, 
in Ref.~\cite{ee500}, part C.
              
\bibitem{ISR} T.~Barklow, P.~Chen and W.~Kozanecki,
in Ref.~\cite{ee500}, part A.

\bibitem{structure} F.A. Berends, W.L. van Neerven and G.J. Burgers,
{\it Nucl. Phys.} {\bf B297} (1988) 429; Erratum, {\it ibidem}
{\bf B304} (1988) 95;\\
E.A. Kuraev and V.S. Fadin, {\it Sov. J. Nucl. Phys.} {\bf 41}
(1985) 466;\\
G. Altarelli and G. Martinelli, Proceedings of the Workshop
`{\it Physics at LEP}', eds. J. Ellis
and R. Peccei, G\`eneva, 1986, CERN 86-02;\\
R. Kleiss, \np B347 1990 29.

\bibitem{Nicro} O.~Nicrosini and L.~Trentadue, {\it Phys. Lett.} {\bf B196}
(1987) 551; {\it Z. Phys.} {\bf C39}  (1988) 479.

\bibitem{BCDKZ} V.~Barger, K.~Cheung, A.~Djouadi, B.A.~Kniehl and P.M.~Zerwas,
                \pr D49 1994 79.

\bibitem{Orange3}  V.~Barger, K.~Cheung, A.~Djouadi, B.A.~Kniehl,
                   R.J.N.~Phillips and
                   P.M.~Zerwas, in  Ref.~\cite{ee500}, part C.

\bibitem{ref7} R.J. Guth and J.H.~K\"uhn, \np B368 1992 38.

\bibitem{4guys} S. Bethke, Z. Kunszt, D.E. Soper and W.J. Stirling,
\np B370 1992 310.

\bibitem{noi5} A. Ballestrero, V.A. Khoze,  E. Maina, S. Moretti
and W.J. Stirling, \preprint\ DFTT 03/95, DTP/95/08, Cavendish--HEP--95/14, 
October 1995 (to be published in {\it Z. Phys.} {\bf C}).

\bibitem{ref81-85} J.F. Donoghue and G. Valencia, {\it Phys. Rev. Lett.}
{\bf 58} (1987) 451; Erratum, {\it ibidem} {\bf 60} (1988) 243;\\ 
W. Bernreuther, U. L\"ow, J.P. Ma and O. Nachtmann, \zp C43 1989 117;\\
W. Bernreuther and O. Nachtmann, \pl B268 1991 424;\\
C.A. Nelson, \pr D41 1990 2805;\\
G.L. Kane, G.A. Ladinsky and C.P. Yuan, \pr D45 1992 124.

\bibitem{ref38} M. Je\.zabek and J.H.~K\"uhn, {\it Phys. Lett.}
{\bf B329} (1994) 317.

\bibitem{ref78} S. Jadach and J.H.~K\"uhn, \preprint\ MPI-PAE/PTh-64/86,
October 1986.

\bibitem{guide} J.F.~Gunion, H.E.~Haber, G.L.~Kane and S.~Dawson,
                ``The Higgs Hunter Guide''
                (Addison-Wesley, Reading MA, 1990).

\bibitem{grosse} P. Grosse-Wiesmann, D.~Haidt and H.J.~Schreiber, in 
Ref.~\cite{ee500}, part A.

\bibitem{preparation} S. Moretti, in preparation.

\end{thebibliography}
\end{document}